\documentclass[12pt]{article}

\usepackage{newtxtext,newtxmath}

\usepackage{graphicx}

\usepackage[letterpaper,margin=1in]{geometry}

\renewenvironment{abstract}
	{\quotation}
	{\endquotation}

\date{}

\makeatletter
\renewcommand{\fnum@figure}{\textbf{Figure \thefigure}}
\renewcommand{\fnum@table}{\textbf{Table \thetable}}
\makeatother

\usepackage{scicite}

\usepackage{url}

\usepackage{soul}
\soulregister{\cite}{7}

\def\scititle{
	 A full-stack analog optical quantum computing platform with one hundred inputs

}
\title{\bfseries \boldmath \scititle}

\author{
	Shota~Yokoyama$^{1\dagger}$,
	Atsushi~Sakaguchi$^{1\dagger}$,
	Warit~Asavanant$^{1,2,3}$,
	Kan~Takase$^{1,2,3}$,\and
	Yi-Ru~Chen$^{1}$,
	Hironari~Nagayoshi$^{2}$,
	Jun-ichi~Yoshikawa$^{1}$,
	Takahiro~Kashiwazaki$^{4}$,\and
	Asuka~Inoue$^{4}$,
	Takeshi~Umeki$^{4}$,
	Toshikazu~Hashimoto$^{4}$,
	Takuji~Hiraoka$^{5}$,\and
	Akira~Furusawa$^{1,2*}$,
	Hidehiro~Yonezawa$^{1*}$,\and
	\small$^{1}$RIKEN Center for Quantum Computing, 2-1 Hirosawa, Wako, Saitama 351-0198, Japan.\and
	\small$^{2}$Department of Applied Physics, School of Engineering, The University of Tokyo, 7-3-1 Hongo,\and
	\small Bunkyo-ku, Tokyo 113-8656, Japan.\and
	\small$^{3}$OptQC Corp., 3-28-13 Nishi-Ikebukuro, Toshima-ku, Tokyo 171-0021, Japan.\and
	\small$^{4}$NTT Device Technology Labs, NTT Corporation, 3-1 Morinosato Wakamiya, Atsugi,\and
	\small Kanagawa 243-0198, Japan.\and
	\small$^{5}$Fixstars Amplify Corporation, 3-1-1 Shibaura, Minato-ku, Tokyo 108-0023, Japan.\and
	\small$^\ast$Corresponding authors. Email: akiraf@ap.t.u-tokyo.ac.jp, hidehiro.yonezawa@riken.jp\and
	\small$^\dagger$These authors contributed equally to this work.
}

\begin{document} 

\maketitle

\begin{abstract} \bfseries \boldmath
	
	Optical technology is a highly promising platform for quantum computing due to its enormous potential for large-scale, ultrafast computation. However, realizing a programmable and scalable system remains a significant challenge. Here, we present a high-speed programmable Gaussian quantum computing platform with one hundred inputs based on a continuous-variable full-stack architecture. Our system features a 100~MHz clock frequency and integrates a cloud-based interface with an open-source Python software development kit, mqc3, significantly enhancing accessibility and operational flexibility. We provide a comprehensive characterization of our system and its capabilities through multi-input and multi-step teleportation, as well as the programmable routing of quantum states across 101 input modes. This platform represents a critical milestone in scalable analog quantum information processing, offering a robust testbed for the future integration of non-Gaussian resources and the development of large-scale optical neural networks.

\end{abstract}

\noindent
Quantum computing is widely regarded as a transformative technology in the field of information technology. By harnessing the principles of quantum mechanics, quantum computers can efficiently solve certain problems that are computationally intractable for modern classical computers~\cite{nielsen2010quantum,gyongyosi2019survey,gill2022quantum}. Quantum computers hold potential for applications including material science~\cite{bauer2020quantum}, finance~\cite{herman2023quantum}, and machine learning~\cite{perdomo2018opportunities}. To date, various physical platforms including superconducting circuits, neutral atoms, trapped ions, silicon-based systems, and photonics have been intensively developed~\cite{ladd2010quantum}. However, most current systems remain in the noisy intermediate-scale quantum (NISQ) era, where practical industrial applications are limited by noise and scalability challenges~\cite{preskill2018quantum,arute2019quantum,zhu2022quantum,kim2023evidence,ai2024quantum,gao2025establishing,zhong2020quantum,aghaee2025scaling}. 

While qubit-based digital architectures dominate the field, analog quantum computing with continuous variables (CVs)~\cite{braunstein2005quantum} offers a compelling alternative for near-term practical utility. Analog computing traditionally excels in low energy consumption and high-speed processing~\cite{maclennan2009analog, maclennan2014promise}, making it particularly suitable for nature-inspired computational models such as neural networks~\cite{krotov2023new,mead1988silicon}. In these models, information is naturally represented by CVs, often tolerating low precision, similar to neural processing in the human brain~\cite{maclennan2009analog}. CV-based analog quantum computing is thus uniquely positioned for tasks involving approximate solutions, large-scale data processing such as in neural networks~\cite{killoran2019continuous,bangar2023experimentally,anand2024time,bangar2025continuous}, and quantum simulation~\cite{kendon2010quantum}.

Optical systems are ideal for this approach due to their capacity for high-bandwidth and large-scale operations~\cite{asavanant2022optical,takeda2019toward,o2007optical}. Specifically, time-domain multiplexing enables the generation of massive entanglement with a compact setup~\cite{yokoyama2013ultra,yoshikawa2016invited,asavanant2019generation,larsen2019deterministic}, providing a scalable platform for measurement-based quantum computing (MBQC)~\cite{raussendorf2001one,menicucci2006universal,briegel2009measurement,menicucci2011temporal}. Despite recent progress, earlier time-domain multiplexed MBQC systems~\cite{asavanant2021time,larsen2021deterministic} were constrained by low clock frequencies (the measurement rate that determines operational speed) and a limited number of input modes: namely, a single input mode at 25~MHz~\cite{asavanant2021time} and six input modes at 4~MHz~\cite{larsen2021deterministic}. These constraints were largely due to the limited bandwidth of the optical parametric oscillators used, which also hindered the implementation of complex, multi-step quantum circuits.

In this paper, we demonstrate a programmable, full-stack analog optical quantum computing platform operating at a 100~MHz clock frequency with a hundred input modes. Here, the clock frequency corresponds to the implementation speed of a two-mode operation in our system. This scale and speed are enabled by our ultra-wideband optical parametric amplifiers (OPAs) featuring a 6~THz bandwidth~\cite{kashiwazaki2023over}. Our system utilizes a time-domain-multiplexed quad-rail lattice cluster state~\cite{alexander2016flexible}, upon which arbitrary multi-mode Gaussian operations are implemented through tailored measurement bases and feedforward operations. 
Accumulated Gaussian noise and spurious biases through multiple computational steps are suppressed by careful calibration and repeated trials without the need for explicit error correction. 
To ensure practicality, we developed mqc3\cite{mqc3SDK}, an open-source Python software development kit (SDK), and an integrated cloud-based interface. These software tools automate the compilation of quantum circuits into hardware parameters, analogous to qubit-based quantum computers~\cite{gill2022quantum}. This architecture serves not only as a near-term testbed but also as a foundation for future fault-tolerant universal quantum computing, as it is compatible with recently reported high-quality Gottesman--Kitaev--Preskill (GKP) qubits~\cite{gottesman2001encoding,konno2024logical,larsen2025integrated}. In our system, fundamental operations on GKP qubits~\cite{menicucci2014fault}, including entangling operations and error syndrome measurements, can be implemented.

We evaluate single-mode and two-mode operations, verifying the quantumness of the process as well as demonstrating multi-input, multi-step operations. In particular, we demonstrate 101-mode, 100-step parallel quantum teleportation, evidencing that unwanted noise and biases introduced through multiple operations can be suppressed and eliminated. In addition, we demonstrate the programmable routing of 101 quantum modes, where input modes with random amplitudes are rearranged in ascending and descending order. These demonstrations validate the reliability, flexibility, and scalability of our system, marking a significant milestone toward practical analog quantum information processing.

\subsection*{Optical quantum computing platform}

Our system utilizes CV electromagnetic field quadratures ($\hat x$, $\hat p$) with a commutation relation of $[\hat x, \hat p] = i$ ($\hbar=1$), where $\hbar$ is the reduced Planck constant~\cite{braunstein2005quantum}.
Figure~\ref{fig1}A illustrates the optical setup as well as the cloud-based system. Figures~\ref{fig1}B and C illustrate the structure of the entanglement created in the optical setup. Four optical wavepackets, referred to as micronodes and labeled as a, b, c, and d, coexisting in time, are grouped as a macronode. The macronodes are interconnected by entanglement, forming a two-dimensional lattice structure on the surface of a cylinder. 
This entanglement has a helical structure, and the number of macronodes $N$ on a single turn of the helix is set to 101 in the current setup, corresponding to one hundred input micronodes, as shown in Figs.~\ref{fig1}C and D. 
Computation is executed through sequential measurements of macronodes, controlled by the measurement angles $\theta^{j}$ ($j=\text{A},\text{B},\text{C}, \text{D}$), followed by feedforward operations, which in the current setup are digitally applied to the measured data~\cite{SupMat}.

In our platform, we implement two-mode macronode operations as illustrated in Figs.~\ref{fig1}D and E. While the system operates through sequential two-mode operations, arbitrary multi-mode Gaussian operations can be realized through proper circuit design. 
 It is well known that arbitrary multi-mode Gaussian operations can be implemented using beam splitter networks and squeezing operations~\cite{braunstein2005squeezing}. The efficient execution of such operations on our system has been established~\cite{yoshikawa2025configurationdesignmultimodegaussian}. Thus, our quantum processor serves as a universal and programmable Gaussian computing platform with one hundred input quantum modes. 

While our quantum computing platform is capable of realizing flexible quantum circuits, manual compilation from arbitrary quantum circuits to graph representations (Figs.~\ref{fig1}C or D) and then to actual machine parameters (the measurement angles) is cumbersome and can be a significant obstacle to testing practical applications. To overcome this difficulty, we have developed an open-source Python SDK, named mqc3~\cite{mqc3SDK}, that enables users to design quantum circuits and automatically compile them into graph representations and subsequent machine parameters. This SDK facilitates the flexible and intuitive implementation of arbitrary Gaussian operations on multi-mode inputs. Furthermore, a cloud-based system has been developed so that users can design quantum circuits remotely using the Python SDK and run the platform. This makes our system accessible to a broader range of users who have a good understanding of problem-solving but may have limited knowledge of the underlying hardware. 

Our analog optical quantum computing platform is specialized for Gaussian operations~\cite{weedbrook2012gaussian} without non-Gaussian functions or error correction capabilities. While it is critical to pursue the extension to the non-Gaussian domain and establish an error correction system, our system offers distinct practical advantages over qubit-based digital platforms. First, it features high processing speeds due to the wide bandwidth of optical devices and the deterministic nature of CV quantum computing~\cite{asavanant2022optical}. Second, it is highly scalable thanks to optical time-domain-multiplexing techniques~\cite{yokoyama2013ultra,yoshikawa2016invited,asavanant2019generation,larsen2019deterministic}. Furthermore, it operates without cryogenic equipment, reducing energy consumption and system complexity.
The analog nature of our system, with a hundred programmable inputs, provides an excellent testbed for applying quantum computing to practical problems, and may open up new applications in fields such as optimization, machine learning, and quantum simulation.

\subsection*{Individual macronode operations}
As a verification of our quantum computing platform, we test several individual macronode operations, which are two-input and two-output operations. These operations are represented by two 50:50 beam splitters and two so-called generalized teleportation operations, as illustrated in Fig.~\ref{fig1}E. The generalized teleportation operations are further decomposed into a series of rotation, squeezing, and rotation operations, which are controlled by the measurement angles $\theta^j$ (see \cite{SupMat} for details). 

A transformation at an individual macronode can be generally expressed as 
\begin{align}
	\boldsymbol{\hat q_\text{out}} = \boldsymbol{S} 
	\boldsymbol{\hat q_\text{in}} + \boldsymbol{\hat \Delta}_{\text{noise}}
	\label{eq:IndivSmat}
\end{align}
where 
$\boldsymbol{\hat q_\text{out}} = (\hat x_{\text{out}1} ,\hat p_{\text{out}1} ,\hat x_{\text{out}2} ,\hat p_{\text{out}2} )^T$ 
is a vector of the output quadratures, 
$\boldsymbol{\hat q_\text{in}} = (\hat x_{\text{in}1} ,\hat p_{\text{in}1} ,\hat x_{\text{in}2} ,\hat p_{\text{in}2} )^T$ 
is a vector of the input quadratures, 
$\boldsymbol{S}=\{ S_{ij}\} \ (i,j=1,\cdots, 4)$ 
is a symplectic transformation matrix, and $\boldsymbol{\hat \Delta}_\text{noise}$ represents the additional Gaussian noise introduced at each operation, which can be suppressed by careful calibration and repeated trials. For small-scale operations, where phase drift in the system is negligible, the variances of the nullifiers (or the squeezing level) are typically 4--5~dB below the shot-noise level. The additional noise level is characterized by $\sqrt{2}$ times these nullifier variances~\cite{SupMat}, while the stability over longer operation sequences is addressed in the subsequent section. Individual macronode operations are executed sequentially to implement multi-input, multi-step computations, as illustrated in Fig.~\ref{fig1}D. The symplectic matrix is estimated through entanglement between the reference modes and the output modes of the operations shown in Fig.~\ref{fig2}A~\cite{SupMat}.

First, we examine a special case of an individual macronode operation: a single-mode operation. By setting the measurement angles $\theta^j$ properly, we can implement identical but independent single-mode operations onto each input at a macronode~\cite{SupMat}. The single-mode operation implementable by a macronode is the generalized teleportation, denoted by $\boldsymbol{V}(\phi_+,\phi_-)$, where
$\phi_\pm = \theta^B \pm \theta^A$ = $\theta^D \pm \theta^C$ 
(consistent sign convention). By adjusting $\phi_+$ and $\phi_-$, we can implement a variety of single-mode operations including teleportation, rotation, squeezing, and shear as illustrated in Fig.~\ref{fig2}B. Figure~\ref{fig2}C shows the theoretical and experimental transformation matrices $\boldsymbol{S}$ as functions of $\phi_+$ and $\phi_-$ with the corresponding formulae. 
Since the same single-mode operations are applied to two input modes, the transformation matrix consists of two identical 2-by-2 submatrices with zero off-diagonal submatrices.
The experimentally reconstructed transformation matrix shows good agreement with the theoretical one over the entire parameter space spanned by $\phi_+$ and $\phi_-$, evidencing the reliable and flexible implementation of a variety of single-mode operations.
The discrepancies in the estimated transformation matrix are evaluated through the normalized Frobenius norm of the difference, $||\boldsymbol{S}_\text{theory} - \boldsymbol{S}_\text{expt} ||/||\boldsymbol{S}_\text{theory} ||$, yielding $4.9 \% \pm 1.1 \%$ averaged over the entire parameter space. It should be emphasized that these deviations primarily stem from uncertainties in the estimation process, not necessarily from the operation itself. Although it is difficult to separate actual operational inaccuracies from estimation-related artifacts, the integrity of the operation is further validated through the multi-step operations in the following sections. 

Next, we examine two-mode operations at individual macronodes. As an example, we demonstrate the generalized controlled-Z operation. This is an entangling operation controlled by two parameters $g$ and $h$, which are determined by the measurement angles $\theta^j$. The theoretical and experimental transformation matrices $\boldsymbol{S}$ are presented in Fig.~\ref{fig2}D, showing good agreement between theory and experiment.
The controllable off-diagonal elements ($S_{23}$ and $S_{41}$) indicate the ability to entangle two input modes. In these figures, the parameters ($g$ and $h$) are swept in the range of $[-2, 2]$ for visualization purposes; however, they are not inherently limited to this range. 
The normalized Frobenius norm of the difference is calculated as $5.2 \% \pm 1.0 \%$ averaged over the parameters spanned by $[-2, 2]$. It is noted again that these discrepancies arise largely from the estimation process.
Other operations, such as beam splitter operations, are shown in \cite{SupMat}.

Apart from the accuracy of the transformation matrices, verifying the non-classical nature of these operations is also critical, as operational noise may render the resulting states classical. To verify this, we measure the entanglement between the reference and output modes after a quantum teleportation operation (i.e., entanglement swapping). The existence of entanglement after the operations ensures the quantum nature of those operations.
Figure~\ref{fig2}E shows the results of two types of teleportation, crossed and twisted quantum teleportation. The normalized correlations, 
$\left \langle \left ( \hat x_{\text{ref}} - \hat x_{\text{out}} \right ) ^2\right \rangle  +\left \langle \left ( \hat p_{\text{ref}} + \hat p_{\text{out}} \right ) ^2\right \rangle$,
are summarized in the tables in the figure. 
The original correlations, which correspond to the nullifier of entanglement or squeezing, are typically $-4$ to $-5$~dB. These correlations are degraded to around $-1.5$~dB due to additional noise at each teleportation step as explained before. However, the correlations after quantum teleportation are still below the corresponding shot-noise levels, proving the existence of entanglement~\cite{asavanant2022optical} and thus confirming the quantum nature of each operational step in our system. These experimental results, combined with the reconstructed transformation matrices, ensure the validity of our platform's basic operations. 

\subsection*{Multi-input multi-step quantum teleportation}

As a demonstration of multi-input, multi-step operations, we perform sequential quantum teleportation. As shown in Fig.~\ref{fig3}A, we utilize 101 input modes for the \textquotedblleft parallel\textquotedblright~teleportation, where each input mode is independently teleported across the two-dimensional lattice from left to right. Figure~\ref{fig3}B shows the results of up to 1,000 steps of this 101-input parallel teleportation. The experiment was repeated for 1,000,000 trials to ensure statistical accuracy.
The input-to-output teleportation gains and the correlations between the reference and output modes are estimated for each input mode as a function of the number of teleportation steps. 

The $x$ and $p$ gains remain at unity up to 1,000 steps, proving that the transformation at each step is reliably executed without accumulated errors. 
The noise powers of the quadrature sums or subtractions between the reference and output modes increase with the number of teleportation operations. This is because each step of the operation adds Gaussian noise. At the initial teleportation step, the noise power is below 0~dB, confirming the existence of quantum entanglement between the reference and the teleported output modes (i.e., entanglement swapping). After the first step, the noise begins accumulating.
Notably, even when the noise power exceeds 0~dB, the levels remain well below the classical benchmark, which is the noise level obtained by an equivalent process performed without squeezing (i.e., using only vacuum states)~\cite{SupMat}. This demonstrates that the noise-reduction advantage provided by the quantum resources persists throughout the 1,000-step operation. 

The experimental noise powers show good agreement with theoretical predictions, although a slight degradation in performance is observed toward 1,000 steps. This behavior is consistent with the evolution of the nullifiers measured over the operational steps~\cite{SupMat}, where small phase drifts during the 1~ms measurement window leads to gradual depletion of entanglement. Nevertheless, the impact of these drifts is relatively minor in the 1,000-step parallel teleportation. It is also observed that the noise power increases linearly with the number of teleportation steps, indicating that the sampling overhead required to achieve a certain precision also scales linearly.

Furthermore, as a complementary characterization to investigate the maximum operational depth of our system, we performed ``helical'' teleportation experiments to track a single mode for an even greater number of steps (the details of which are provided in~\cite{SupMat}). Our results reveal that the noise power remains consistent with the theoretical model for up to approximately 10,000 steps. Beyond this extensive range, excess noise emerges due to the depletion of entanglement, and the accumulation of noise and minor technical offsets begins to limit the precision of gain estimation. Defining this practical operational envelope, spanning four orders of magnitude in step count, demonstrates the exceptional stability of our platform and its readiness for executing long-sequence quantum processes. Overall, these results confirm that our platform can reliably manage one hundred inputs and execute a large number of operational steps.

\subsection*{Programmable routing of quantum states}

The large-scale, flexible architecture of our platform enables the implementation of complex, programmable quantum circuits. One of the most critical functions of this architecture is the programmable routing of quantum states. In our setup, external quantum states can be coupled into the platform via optical switches that inject external quantum states into designated input modes (refer to~\cite{SupMat} for the interface design). This interface allows the platform to incorporate non-Gaussian resource states—such as photon-number states, cubic phase states, or GKP qubits—which are essential for achieving a quantum advantage. However, because such resources are typically generated probabilistically~\cite{konno2024logical,larsen2025integrated}, they appear at random input ports (i.e., unpredictable time slots), as illustrated in Fig.~\ref{fig4}A.
To utilize these stochastic resources effectively, our platform enables the on-demand reconfiguration of the quantum circuit. By selectively combining crossed teleportation (which propagates the state straight through the graph) and twisted teleportation (which directs the state at a right angle), we can programmatically route any input to a desired output mode. This high degree of reconfigurability allows for the consolidation of randomly distributed resources into specific target modes, enabling subsequent deterministic multi-mode operations.

To verify this functionality, we demonstrate the programmable routing of 101 input modes based on their displacement amplitudes. In this demonstration, we prepare input modes with random displacements and design a quantum circuit to route them into specific output ports in a designated order. The quantum circuits are implemented via our Python SDK, and the corresponding graph representations are provided in~\cite{SupMat}.
Figure~\ref{fig4}B presents the experimental results for the 101-mode configuration. Despite the initial modes being randomly displaced, they are accurately routed to the target output ports according to their amplitudes in both ascending and descending orders. The successful execution of this programmable routing for 101 randomly displaced inputs confirms the reliable operation of complex, large-scale circuits, proving the versatility of our platform for future large-scale quantum information processing. 

\subsection*{Discussion and Conclusion}

We have developed a large-scale analog optical quantum computing platform capable of executing programmable Gaussian operations on one hundred input modes at a 100~MHz clock frequency. By leveraging time-domain multiplexing and a Python SDK, we have successfully verified fundamental single- and two-mode operations, multi-step teleportation, and programmable routing. 
Since these operations constitute a universal set for Gaussian quantum computation~\cite{yoshikawa2025configurationdesignmultimodegaussian}, our results confirm that the platform is capable of implementing arbitrary multi-mode Gaussian transformations. This exceptional flexibility and scalability provide a robust foundation for complex quantum information processing and future universal quantum computing.

Our experiments show that the Gaussian noise increases linearly with the number of operational steps. This indicates that the sampling overhead for error mitigation remains at a manageable scale as the circuit depth increases. Increasing the initial squeezing level would further reduce this overhead and prolong the preservation of the state's quantum nature, which is essential for the future integration of non-Gaussian resources. While our current system utilizes $-4$ to $-5$~dB of effective squeezing, $-12$~dB squeezing has been reported using a similar broadband waveguide OPA~\cite{ha2026generation12dbsqueezed}. Existing technologies thus provide a clear pathway to higher-performance operations.

Further pursuing scalability is essential, as higher operational speeds significantly enhance total data throughput and accommodate the massive resource overhead required for future fault-tolerant quantum computing. Specifically, the clock frequency and the number of input modes can be significantly extended beyond our current demonstration. While the clock frequency is currently limited to 100~MHz by the bandwidth of the electronic circuits, recent advancements in real-time quadrature measurements of squeezed states up to 43~GHz~\cite{inoue2023toward} and entanglement up to 25~GHz~\cite{kawasaki2025real} suggest that 10~GHz clock frequency is within reach.
Such an increase would reduce the wavepacket duration to a hundredth of its current value, potentially enabling the handling of up to 10,000 inputs within the same optical setup. 
It is noteworthy that even at a 10~GHz clock with 100~ps pulses, fiber dispersion over our current 200~m delay line remains negligible, preserving the temporal mode profiles without complex compensation.
To support deeper circuits with more computational steps, the measurement window—which is currently limited to 1~ms within each 1.7~ms cycle due to the 0.7~ms required for phase locking—could be extended. This limitation can be overcome by upgrading to a locking system utilizing auxiliary beams, which allow for continuous phase control during measurement without interference with the quantum modes.

Furthermore, by harnessing all-optical techniques~\cite{yamashima2025all} to bypass electronic feedforward, it might be possible to exploit the full 6~THz bandwidth of our optical parametric amplifiers. This broad bandwidth also opens the door to integrating frequency-domain multiplexing into our current architecture. While frequency-domain multiplexing~\cite{chen2014experimental,menicucci2008one} alone may not match the scale of entanglement generation and measurement typically achieved with time-domain techniques, it provides the critical advantage of simultaneous multi-mode access. By combining these approaches, one could leverage the high resource efficiency of time-domain multiplexing alongside the parallel processing capabilities of the frequency domain, potentially enabling an even higher degree of multiplexing and more sophisticated large-scale quantum operations.
Such scaling is not only vital for reaching millions of modes but also enhances operational flexibility; for instance, a larger number of available time slots could facilitate the efficient management of probabilistically generated non-Gaussian resources and support high-dimensional applications.

Realizing universal, non-Gaussian quantum computing remains our primary objective. As discussed in~\cite{SupMat}, externally generated non-Gaussian states can be coupled to our platform through a high-speed optical switch. Non-Gaussian operations can then be implemented by incorporating these ancillary states with high-speed, non-linear feedforward~\cite{gottesman2001encoding,sakaguchi2023nonlinear}. The programmable routing functionality demonstrated in this work will be useful for this purpose, as it allows for the reordering of stochastic resource states. The next critical milestone is the implementation of real-time non-linear feedforward to enable non-Gaussian transformations. Our platform serves as an ideal testbed for investigating the potential of analog quantum computing, providing a versatile environment to explore new frontiers in quantum neural networks, optimization, and fault-tolerant quantum information technology.



\begin{figure} 
	\centering
	\includegraphics[width=0.95\textwidth]{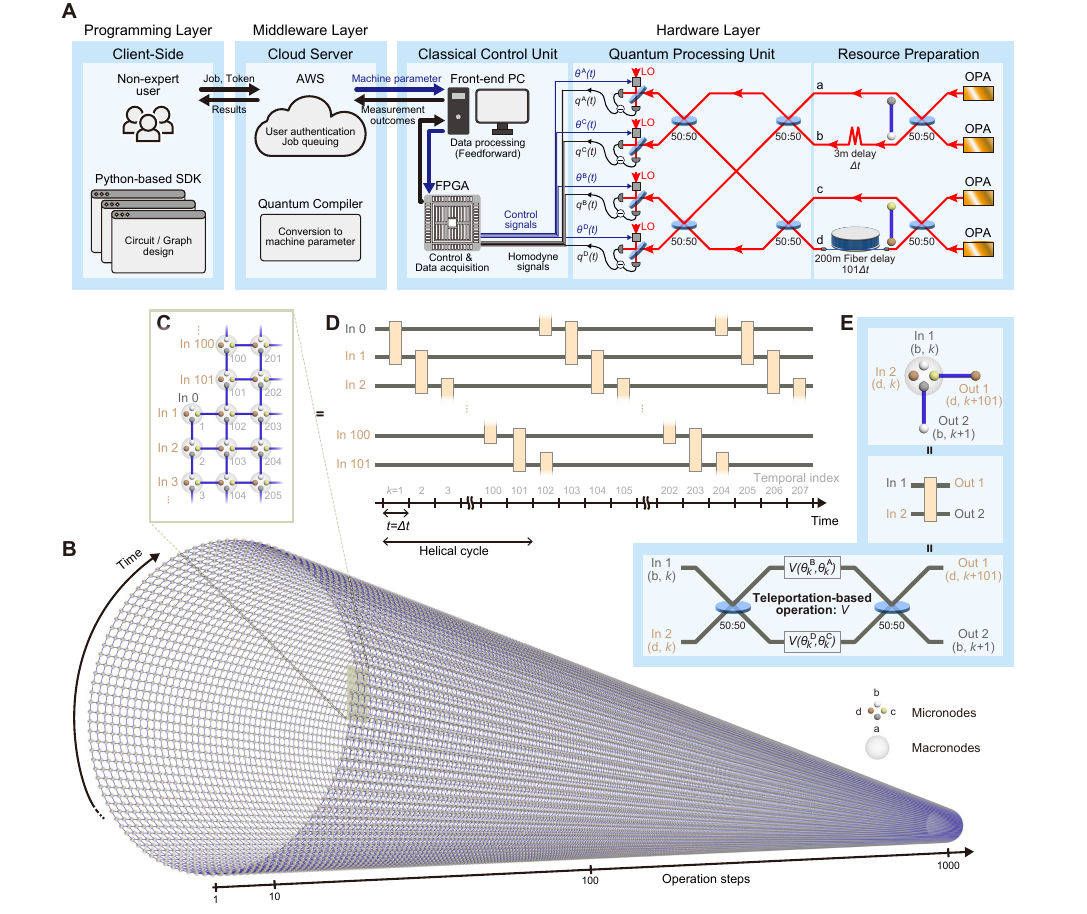}
	
	\caption{\textbf{Cloud-based optical quantum computing platform.} 
		({\bf A}) Schematic of the system including the programming layer (client-side), the middleware layer (cloud server on Amazon Web Services, AWS), and the hardware layer (classical control unit, quantum processing unit, and resource preparation). FPGA: Field Programmable Gate Array. LO: Local Oscillator. OPA: Optical Parametric Amplifier.
		({\bf B}) Three-dimensional representation of the experimentally generated entanglement in the time domain. 
		({\bf C}) Two-dimensional representation of the generated entanglement. One hundred inputs at the edges are processed through the entanglement. 
		({\bf D}) Quantum circuit representation of our system, consisting of sequential two-input, two-output macronode operations. It is noted that arbitrary multi-mode Gaussian operations can be realized by a sequence of these macronode operations. 
		({\bf E}) Operation at each macronode. At the $k$-th macronode, the two inputs are micronodes b and d, and the two outputs are micronode b at $(k+1)$-th macronode and micronode d at $(k+N)$-th macronode. The operation is represented by generalized quantum teleportation sandwiched between two 50:50 beam splitters.}
	\label{fig1}
\end{figure}

\begin{figure}
	\includegraphics[width=0.95\textwidth]{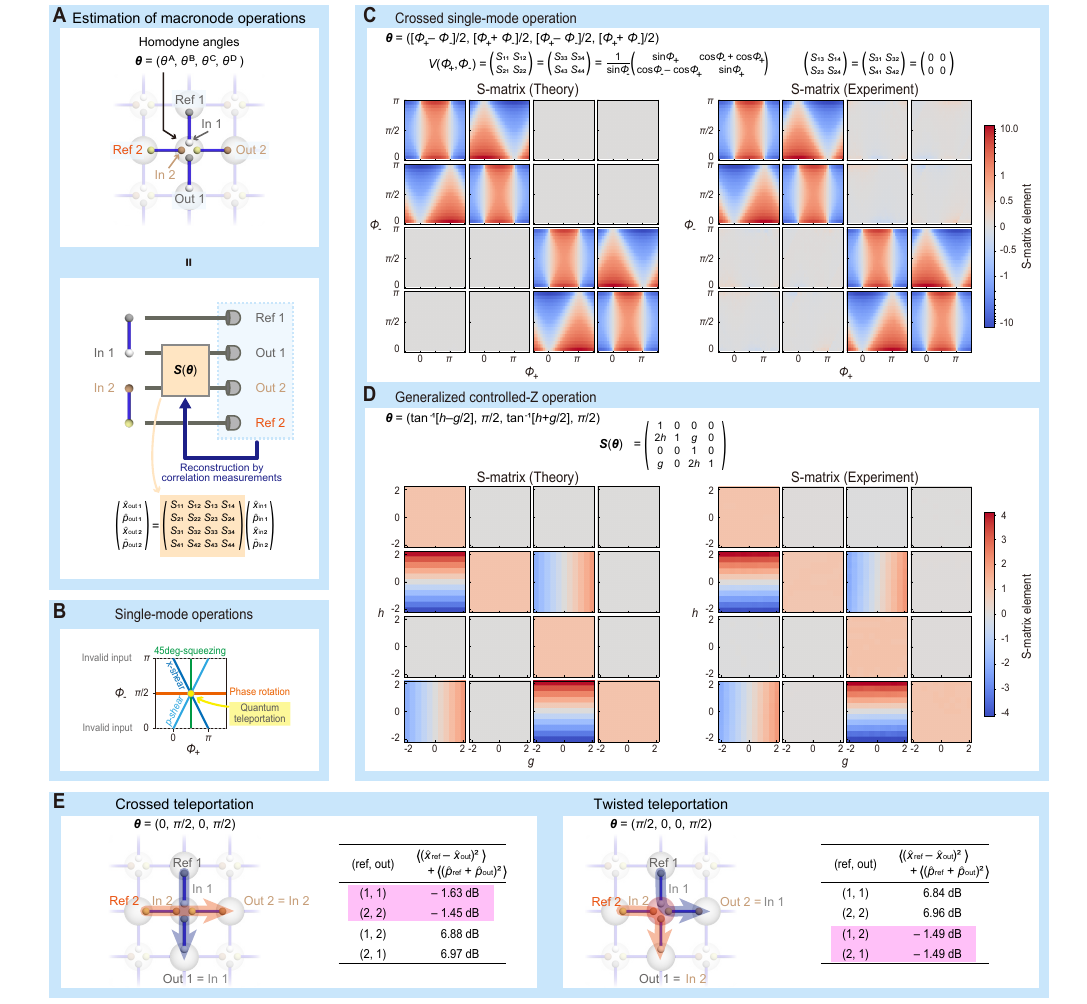}
	\caption{\textbf{Examples of individual macronode operations.}
		({\bf A}) Estimation procedure for the transformation matrix $\boldsymbol{S}$. Two reference modes and two output modes are measured, and then the transformation matrix is estimated through the correlations among them~\cite{SupMat}.
		({\bf B}) Single-mode operations implementable at an individual macronode. The operation is controlled by two parameters, $\phi_+$ and $\phi_-$, which are determined by the measurement angles $\theta^j$.
		Quantum teleportation, phase rotation, squeezing with 45 degrees rotation, and $x$- and $p$-shear operations are illustrated.
		({\bf C}) Transformation matrices for crossed single-mode quantum operations.
		Left: Theoretical $\boldsymbol{S}$ matrix.
		Right: Experimental $\boldsymbol{S}$ matrix. 
		The matrix elements are plotted as two-dimensional color maps spanned by $\phi_+$ and $\phi_-$.
		({\bf D}) Generalized controlled-Z operation. 
		Left: Theoretical $\boldsymbol{S}$ matrix.
		Right: Experimental $\boldsymbol{S}$ matrix. The matrix elements are plotted as two-dimensional color maps spanned by $g$ and $h$. 
		({\bf E}) Verification of quantum nature after quantum teleportation. Left: Crossed teleportation. Right: Twisted teleportation.
	}\label{fig2}
\end{figure}

\begin{figure}
	\includegraphics[width=1\textwidth]{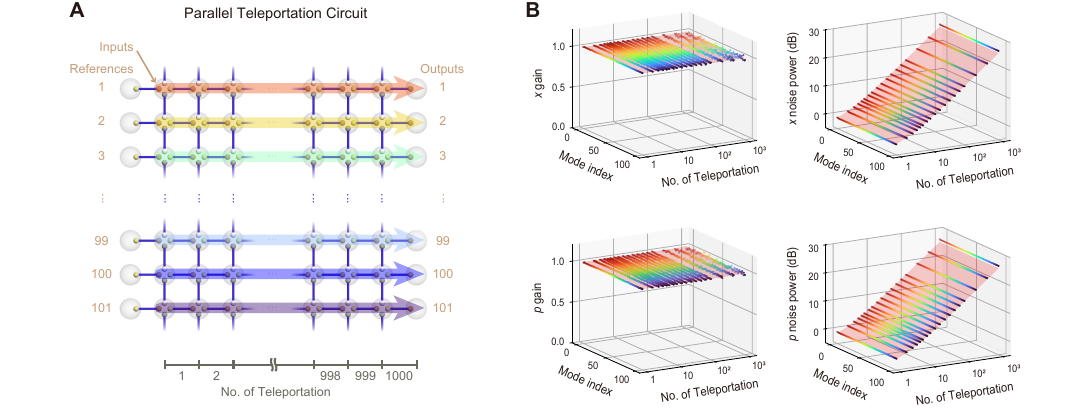}
	\caption{\textbf{Multi-input, multi-step quantum teleportation.} 
		({\bf A}) Graph representation of parallel teleportation. 
		({\bf B}) Results of parallel teleportation. Top two panels show the $\hat x$ quadratures and the bottom two panels show the $\hat p$ quadratures. The left panels represent the teleportation gains, while the right panels represent the noise powers of the quadrature sums for $\hat p$ or subtractions for $\hat x$ between the reference and the output modes after teleportation. The semi-transparent surfaces represent the theoretical predictions. }
	\label{fig3}
\end{figure}

\begin{figure}
	\includegraphics[width=1\textwidth]{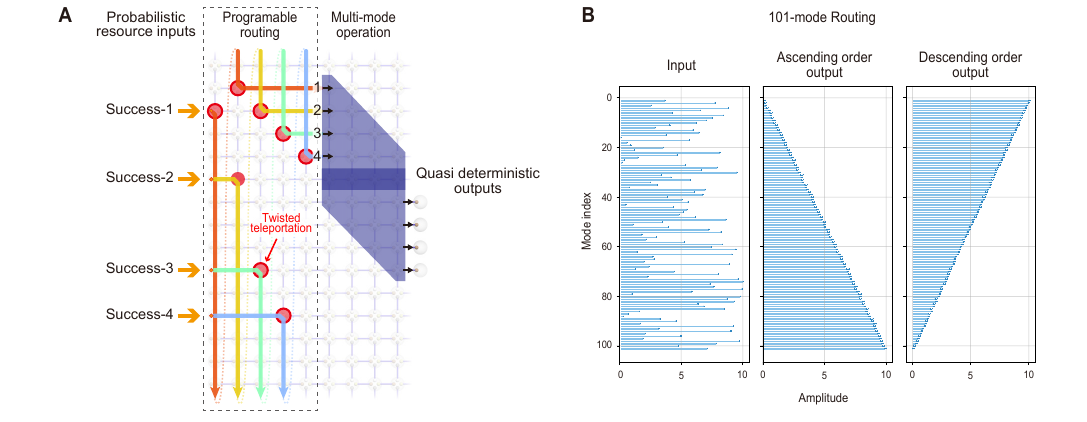}
	\caption{\textbf{Programmable routing of quantum states.} 
		({\bf A}) Conceptual illustration of programmable routing for managing probabilistically generated resources. Non-Gaussian resource states, generated at random positions (time slots), are routed and consolidated into specific modes to facilitate subsequent multi-mode operations. 
		({\bf B}) Experimental demonstration of 101-mode quantum state routing. Input quantum states are randomly displaced and subsequently routed into a designated order according to their amplitudes.}
	\label{fig4}
\end{figure}


\clearpage 

%
\bibliography{AOQC_Bib} 
\bibliographystyle{sciencemag}

%
%
%
%
%
%


\section*{Acknowledgments}
We thank Professor Kenji Doya for valuable comments on the application of optical quantum computing.

\paragraph*{Funding:}
This work was supported by the Japan Science and Technology (JST) Agency (Moonshot R \& D) Grant No. JPMJMS2064 and JPMJMS256I, the UTokyo Foundation, and donations from Nichia Corporation of Japan. 
H.N. acknowledges financial support from The Forefront Physics and Mathematics Program to Drive Transformation (FoPM). H.N. and W.A. acknowledge funding from the Japan Society for the Promotion of Science KAKENHI (No. 23K13040, 24KJ0745).

\paragraph*{Author contributions:}
A.F. and H.Y. led and supervised the project. S.Y., A.S. and H.Y. designed and built the experimental setup with assistance from W.A., K.T. and J.Y. A.S. designed and built the FPGA and electronic systems. Y.C. adjusted and modified the experimental setup. S.Y. prepared the code, recorded and analyzed the data with assistance from A.S. and H.Y. 
W.A., H.N. and S.Y. performed basic theoretical analysis and calculations necessary for SDK and cloud system.
W.A., H.N., K.T., A.S., S.Y., H.Y., and J.Y. discussed theoretical analysis, SDK, and the cloud system. T.~Hiraoka conceived and supervised the development of the SDK, cloud system and related software. T.K., A.I., T.U. and T.~Hashimoto developed and provided the OPAs used in the experiment. H.Y. and S.Y. wrote the paper with input from all authors. 

\paragraph*{Competing interests:}
A.F. and H.Y. are co-founders and directors of OptQC Corp. The authors declare no other competing interests.

\paragraph*{Data and materials availability:}
Additional data are available in the supplementary materials.

\subsection*{Supplementary materials}
Materials and Methods\\
Figs. S1 to S19\\
References \textit{(\arabic{enumiv})}\\ 


\newpage


\renewcommand{\thefigure}{S\arabic{figure}}
\renewcommand{\thetable}{S\arabic{table}}
\renewcommand{\theequation}{S\arabic{equation}}
\renewcommand{\thepage}{S\arabic{page}}
\setcounter{figure}{0}
\setcounter{table}{0}
\setcounter{equation}{0}
\setcounter{page}{1} 


\begin{center}
\section*{Supplementary Materials for\\ \scititle}

Shota~Yokoyama$^{\dagger}$,
Atsushi~Sakaguchi$^{\dagger}$,
Warit~Asavanant,
Kan~Takase,\and
Yi-Ru~Chen,
Hironari~Nagayoshi,
Jun-ichi~Yoshikawa,
Takahiro~Kashiwazaki,
Asuka~Inoue,
Takeshi~Umeki,
Toshikazu~Hashimoto,
Takuji~Hiraoka,
Akira~Furusawa$^{*}$,
Hidehiro~Yonezawa$^{*}$\\

\small$^\ast$Corresponding authors. Email: akiraf@ap.t.u-tokyo.ac.jp, hidehiro.yonezawa@riken.jp\\
\small$^\dagger$These authors contributed equally to this work.

\end{center}

\subsubsection*{This PDF file includes:}
Materials and Methods\\
Figures S1 to S19\\


\newpage


\section*{Materials and Methods}

\section*{S1 \ Measurement-based quantum computation using quad-rail lattice cluster state}

\subsection*{S1.1 Quantum resource for measurement-based quantum computation}

In this section, we describe the structure of the quad-rail lattice (QRL) cluster state\cite{menicucci2011temporal,alexander2016flexible,walshe2021streamlined}, which serves as a resource for measurement-based quantum computation. We also provide equivalent representations that will be used for further calculations in the subsequent sections.

\begin{figure} 
	\centering
	\includegraphics[width=0.8\textwidth]{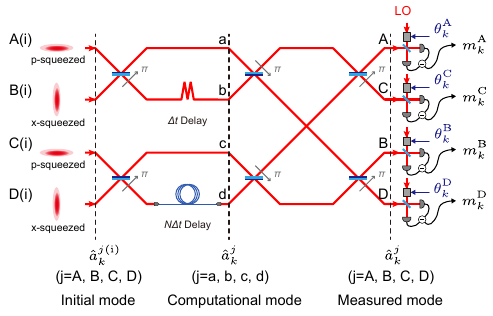}
	
	\caption{\textbf{Schematic of measurement-based quantum computation using quad-rail lattice cluster states.}}
	\label{figS1} 
\end{figure}

Figure~\ref{figS1} shows the schematics of our optical setup. 
Four squeezed vacuum states, which are generated by optical parametric amplifiers (OPAs), are the initial modes of this setup, whose amplitude operators are given by, 
\begin{align}
	\sqrt{2} \hat a^{j(\text{i})}_{k}  = 
	\begin{cases}
		\mathrm{e}^{r_j} \hat x^{j(0)}_k + i \mathrm{e}^{-r_j} \hat p^{j(0)}_k & (j = \text{A},\text{C}) \\
		\mathrm{e}^{-r_j} \hat x^{j(0)}_k + i \mathrm{e}^{r_j} \hat p^{j(0)}_k & (j = \text{B},\text{D}). \\
	\end{cases}
\end{align}
where $r_j \left( \geq 0 \right)$ is a squeezing parameter and $k$ is the time step or the index of wavepackets. 
Modes A and C correspond to $\hat{p}$-squeezed vacuum states, while modes B and D correspond to $\hat{x}$-squeezed vacuum states. Here, $\hat q^{j(0)}_k$ $(\hat q = \hat x,  \hat p)$ denotes the quadrature operators of vacuum modes, which satisfy:
\begin{align}
	\big \langle \hat q_k^{j(0)} \big\rangle = 0, \qquad 
	\big \langle \big( \hat q_k^{j(0)} \big)^2 \big\rangle = \frac{1}{2},
\end{align}
where $\hbar=1$. 

The initial squeezed vacuum modes first pass through two 50:50 beam splitters, generating two two-mode entangled states or so-called Einstein-Podolsky-Rosen (EPR) states. Then two optical delay lines with time delays of $\Delta t$ and $N \Delta t$ are applied, resulting in computational modes (also referred to as distributed modes) labeled as a, b, c, and d as shown in Fig.~\ref{figS1}. Since the time duration of a wavepacket is $\Delta t$, the $\Delta t$-delay line shifts the wavepacket by one wavepacket (or one time step), while the $N\Delta t$-delay line shifts the wavepacket by $N$ wavepackets (or $N$ time steps). Note that a wave packet is also referred as a micronode in a graph representation, and four micronodes in the same time slot are grouped as a macronode.

\begin{figure}
	\centering
	\includegraphics[width=0.5\textwidth]{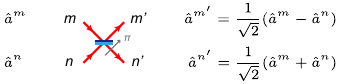}
	\caption{\textbf{Beam splitter transformation in the Heisenberg picture used in this Supplementary Materials.}}
	\label{figS2}
\end{figure}

The definition of a beam splitter may vary slightly across the literature. In this Supplementary Materials, we define the transformation of a 50:50 beam splitter between modes $m$ and $n$ in the Heisenberg picture as shown in Fig.~\ref{figS2}:
\begin{align}
	\begin{pmatrix}
		\hat a ^{m'} \\
		\hat a ^{n'} 
	\end{pmatrix}
	=
	\hat B^\dagger _{m,n} 
	\begin{pmatrix}
		\hat a ^{m} \\
		\hat a ^{n} 
	\end{pmatrix}
	\hat B_{m,n}
	=
	\boldsymbol{B}_0
	\begin{pmatrix}
		\hat a ^{m} \\
		\hat a ^{n} 
	\end{pmatrix}
	, \qquad 
	\boldsymbol{B}_0 = \frac{1}{\sqrt{2}}
	\begin{pmatrix}
		1 & -1 \\
		1 & 1
	\end{pmatrix}.
\end{align}

From the above transformations and the time-step shifts introduced by the optical delay lines, we find that the computational modes can be expressed in terms of the initial modes as
\begin{align}
	\begin{pmatrix}
		\hat q_k ^\text{a} \\
		\hat q_k ^\text{b} \\
		\hat q_k ^\text{c} \\
		\hat q_k ^\text{d} \\
	\end{pmatrix} 
	= \frac{1}{\sqrt{2}}
	\begin{pmatrix}
		\hat q_k ^{\text{A(i)}} - \hat q_k ^{\text{B(i)}} \\
		\hat q_{k-1} ^{\text{A(i)}} + \hat q_{k-1} ^{\text{B(i)}} \\
		\hat q_k ^{\text{C(i)}} - \hat q_k ^{\text{D(i)}} \\
		\hat q_{k-N} ^{\text{C(i)}} + \hat q_{k-N} ^{\text{D(i)}} \\
	\end{pmatrix}.
\end{align}
This implies that $k$-th macronode in the computational modes are entangled with $(k-1)$-th, $(k+1)$-th, $(k-N)$-th and $(k+N)$-th macronodes through the distributed EPR pairs. The structure of entanglement is shown in Fig.\ref{fig1}B of the main text.

Next, these computational modes pass through four beam splitters (also referred to as a four-splitter), resulting in the measured modes A, B, C, and D in Fig.~\ref{figS1}. The relationship between the measured modes A, B, C, and D, and the computational modes a, b, c, and d can be described by the transformation of a four-splitter:
\begin{align}
	\begin{pmatrix}
		\hat q_k ^\text{A} \\
		\hat q_k ^\text{B} \\
		\hat q_k ^\text{C} \\
		\hat q_k ^\text{D} \\
	\end{pmatrix}
	= \boldsymbol{B}
	\begin{pmatrix}
		\hat q_k ^\text{a} \\
		\hat q_k ^\text{b} \\
		\hat q_k ^\text{c} \\
		\hat q_k ^\text{d} \\
	\end{pmatrix},
	\qquad 
	\boldsymbol{B} = (\boldsymbol{B}_0\otimes \boldsymbol{I}_2)(\boldsymbol{I}_2\otimes \boldsymbol{B}_0^{-1})
	= 
	\frac{1}{2}
	\begin{pmatrix}
		1 & 1 & -1 & -1\\
		-1 & 1 & 1 & -1\\
		1 & 1 & 1 & 1\\
		-1 & 1 & -1 & 1
	\end{pmatrix},
	\label{eq:foursplitter}
\end{align}
where $\boldsymbol{I}_2$ denotes 2 by 2 identity matrix.
As a result, the measured quadratures at the homodyne detectors can be expressed in terms of the quadratures of initial modes\cite{yokoyama2013ultra,asavanant2019generation}:
\begin{align}
	\begin{pmatrix}
		\hat q_k ^\text{A} \\
		\hat q_k ^\text{B} \\
		\hat q_k ^\text{C} \\
		\hat q_k ^\text{D} \\
	\end{pmatrix}
	=
	\frac{1}{2\sqrt{2}}
	\begin{pmatrix}
		\phantom{+} \hat q_k ^{\text{A(i)}} - \hat q_k ^{\text{B(i)}} - \hat q_k ^{\text{C(i)}} + \hat q_k ^{\text{D(i)}} + \hat q_{k-1} ^{\text{A(i)}} + \hat q_{k-1} ^{\text{B(i)}} - \hat q_{k-N} ^{\text{C(i)}} - \hat q_{k-N} ^{\text{D(i)}} \\
		- \hat q_k ^{\text{A(i)}} + \hat q_k ^{\text{B(i)}} + \hat q_k ^{\text{C(i)}} - \hat q_k ^{\text{D(i)}} + \hat q_{k-1} ^{\text{A(i)}} + \hat q_{k-1} ^{\text{B(i)}} - \hat q_{k-N} ^{\text{C(i)}} - \hat q_{k-N} ^{\text{D(i)}} \\
		\phantom{+} \hat q_k ^{\text{A(i)}} - \hat q_k ^{\text{B(i)}} + \hat q_k ^{\text{C(i)}} - \hat q_k ^{\text{D(i)}} + \hat q_{k-1} ^{\text{A(i)}} + \hat q_{k-1} ^{\text{B(i)}} + \hat q_{k-N} ^{\text{C(i)}} + \hat q_{k-N} ^{\text{D(i)}} \\
		- \hat q_k ^{\text{A(i)}} + \hat q_k ^{\text{B(i)}} - \hat q_k ^{\text{C(i)}} + \hat q_k ^{\text{D(i)}} + \hat q_{k-1} ^{\text{A(i)}} + \hat q_{k-1} ^{\text{B(i)}} + \hat q_{k-N} ^{\text{C(i)}} + \hat q_{k-N} ^{\text{D(i)}} \\
	\end{pmatrix}.	
\end{align}

From these relationships, we derive the equivalent \textit{nullifier representations}, specifically those for the quad-rail lattice cluster state, EPR pairs, and the underlying squeezed states:
\begin{align}
	\begin{pmatrix}
		\hat \delta_k ^\text{A} \\
		\hat \delta_k ^\text{B} \\
		\hat \delta_k ^\text{C} \\
		\hat \delta_k ^\text{D} \\
	\end{pmatrix}
	& \equiv
	\frac{1}{2\sqrt{2}}
	\begin{pmatrix}
		\phantom{-} \hat p_k ^\text{A} - \hat p_k ^\text{B} + \hat p_k ^\text{C} - \hat p_k ^\text{D} + \hat p_{k+1} ^\text{A} + \hat p_{k+1} ^\text{B} + \hat p_{k+1} ^\text{C} + \hat p_{k+1} ^\text{D}\\
		- \hat x_k ^\text{A} + \hat x_k ^\text{B} - \hat x_k ^\text{C} + \hat x_k ^\text{D} + \hat x_{k+1} ^\text{A} + \hat x_{k+1} ^\text{B} + \hat x_{k+1} ^\text{C} + \hat x_{k+1} ^\text{D}\\
		- \hat p_k ^\text{A} + \hat p_k ^\text{B} + \hat p_k ^\text{C} - \hat p_k ^\text{D} - \hat p_{k+N} ^\text{A} - \hat p_{k+N} ^\text{B} + \hat p_{k+N} ^\text{C} + \hat p_{k+N} ^\text{D}\\
		\phantom{-} \hat x_k ^\text{A} - \hat x_k ^\text{B} - \hat x_k ^\text{C} + \hat x_k ^\text{D} - \hat x_{k+N} ^\text{A} - \hat x_{k+N} ^\text{B} + \hat x_{k+N} ^\text{C} + \hat x_{k+N} ^\text{D}\\
	\end{pmatrix} 
	\label{eq:HD_nullifier} \\
	& = \frac{1}{\sqrt{2}}
	\begin{pmatrix}
		\phantom{-} \hat p_k ^\text{a} + \hat p_{k+1} ^\text{b} \\
		- \hat x_k ^\text{a} + \hat x_{k+1} ^\text{b} \\
		\phantom{-} \hat p_k ^\text{c} + \hat p_{k+N} ^\text{d} \\
		- \hat x_k ^\text{c} + \hat x_{k+N} ^\text{d} \\
	\end{pmatrix}	
	\label{eq:EPR_nullifier} \\
	& =
	\begin{pmatrix}
		\hat p_k ^\text{A(i)} \\
		\hat x_k ^\text{B(i)} \\
		\hat p_k ^\text{C(i)} \\
		\hat x_k ^\text{D(i)} \\
	\end{pmatrix} 
	\label{eq:Squeeze_nullifier} \\
	& =
	\begin{pmatrix}
		e^{-r_\text{A}}  \hat p_k ^\text{A(0)} \\
		e^{-r_\text{B}}  \hat x_k ^\text{B(0)} \\
		e^{-r_\text{C}}  \hat p_k ^\text{C(0)} \\
		e^{-r_\text{D}}  \hat x_k ^\text{D(0)} \\
	\end{pmatrix}.
	\label{eq:SqueezePara_nullifier} 
\end{align}

\subsection*{S1.2 \ Measurement and numerical Feedforward}
Quantum operations are implemented by repeatedly measuring the quadratures at each time step (or macronode) $k$, denoted as: 
\begin{align}
	\hat m_k^j = \hat x_k^{j} \sin \theta^j_k + \hat p_k^{j} \cos \theta^j_k \ \ (j = \text{A}, \text{B}, \text{C}, \text{D}).
	\label{eq:m}
\end{align}
At macronode $k$, the computational modes b and d are the input modes, which are processed and output in the computational mode b at macronode $k+1$, and mode d at macronode $k+N$. The values of $\theta^j_k$ in Eq.~\eqref{eq:m} are chosen based on the target quantum operations described in later sections. 

The measurement outcomes of those quadratures, $\boldsymbol{m}_k = (m_k^\text{A}, m_k^\text{B}, m_k^\text{C}, m_k^\text{D})^T$, are used to perform feedforward displacements which are applied to the computational modes $(\text{b}, k+1)$ and $(\text{d}, k+N)$: 
\begin{align}
	\begin{pmatrix}
		\hat x_{k+1}^\text{b} \\
		\hat p_{k+1}^\text{b} \\
		\hat x_{k+N}^\text{d} \\
		\hat p_{k+N}^\text{d} \\
	\end{pmatrix}
	\xrightarrow[\text{displacement}]{\text{feedforward}}
	\begin{pmatrix}
		\hat x_{k+1}^\text{b} \\
		\hat p_{k+1}^\text{b} \\
		\hat x_{k+N}^\text{d} \\
		\hat p_{k+N}^\text{d} \\
	\end{pmatrix}
	+
	\boldsymbol{E}_k
	\boldsymbol{m}_k
	+
	\begin{pmatrix}
		\text{x}_{k+1}^\text{b} \\
		\text{p}_{k+1}^\text{b} \\
		\text{x}_{k+N}^\text{d} \\
		\text{p}_{k+N}^\text{d} \\
	\end{pmatrix},
	\label{eq:FFDisp}
\end{align}
where $\boldsymbol{E}_k(\boldsymbol{\theta}_k)$ is a feedforward coefficient matrix determined by the measurement angles $\boldsymbol{\theta}_k =\left( \theta_k^\text{A}, \theta_k^\text{B}, \theta_k^\text{C}, \theta_k^\text{D} \right)^T $ at macronode $k$. The terms $\text{x}_{k+1}^\text{b}$ and $\text{p}_{k+1}^\text{b}$ [$\text{x}_{k+N}^\text{d}$ and $\text{p}_{k+N}^\text{d}$] represent additional displacements applicable to the mode $(\text{b}, k+1)$ [$(\text{d}, k+N)$] that are independent of the feedforward.

In the experiment, all modes, including those to which feedforward is applied, are eventually measured via homodyne detections. Since the four-splitter operation and the homodyne detection that should follow the feedforward displacement are linear, the feedforward displacement can be effectively implemented numerically after the measurement outcomes are acquired.

The supposed measurement outcomes $\boldsymbol{m}_k$ at macronode $k$ with feedforward displacements are numerically calculated by adding the preceding measurement outcomes at macronodes $k-1$ and $k-N$ ($\boldsymbol{m}_{k-1}$ and $\boldsymbol{m}_{k-N}$), and arbitrary displacements 
$\left( \text{x}_{k}^\text{b}, \text{p}_{k}^\text{b}, \text{x}_{k}^\text{d}, \text{p}_{k}^\text{d} \right)^T$ 
to the raw measured outcomes $\boldsymbol{m}'_k$ (i.e., without feedforward operation):
\begin{align}
	\boldsymbol{m}_k
	=
	\boldsymbol{m}'_k
	&+
	\boldsymbol{F}_k 	\boldsymbol{m}_{k-1}
	+
	\boldsymbol{G}_{k}	\boldsymbol{m}_{k-N}
	+
	\boldsymbol{H}_k	
	\begin{pmatrix}
		\text{x}_{k}^\text{b} \\
		\text{p}_{k}^\text{b} \\
		\text{x}_{k}^\text{d} \\
		\text{p}_{k}^\text{d} \\
	\end{pmatrix},
	\label{eq:NumFF}
\end{align}
where the matrix $\boldsymbol{F}_{k}(\boldsymbol{\theta}_k, \boldsymbol{\theta}_{k-1})$ corresponds to a feedforward coefficient matrix from $k-1$ to $k$, $\boldsymbol{G}_{k}(\boldsymbol{\theta}_k, \boldsymbol{\theta}_{k-N})$ represents that from $k-N$ to $k$, and $\boldsymbol{H}_k (\boldsymbol{\theta}_k)$ is a coefficient matrix for the arbitrary displacement.	
These feedforward matrices can be explicitly written as:
\begin{align}
	\boldsymbol{F}_k
	&=
	\left[
	\text{diag}(\sin \boldsymbol{\theta}_k)
	\boldsymbol{B}
	\boldsymbol{e}_2
	\boldsymbol{e}_1^T
	+
	\text{diag}(\cos \boldsymbol{\theta}_k)
	\boldsymbol{B}
	\boldsymbol{e}_2
	\boldsymbol{e}_2^T
	\right]
	\boldsymbol{E}_{k-1}, 
	\\
	\boldsymbol{G}_k
	&=
	\left[
	\text{diag}(\sin \boldsymbol{\theta}_k)
	\boldsymbol{B}
	\boldsymbol{e}_4
	\boldsymbol{e}_3^T
	+
	\text{diag}(\cos \boldsymbol{\theta}_k)
	\boldsymbol{B}
	\boldsymbol{e}_4
	\boldsymbol{e}_4^T
	\right]
	\boldsymbol{E}_{k-N}, 
	\\
	\boldsymbol{H}_k	
	& = 
	\text{diag}(\sin \boldsymbol{\theta}_k)
	\boldsymbol{B}\left(
	\boldsymbol{e}_2
	\boldsymbol{e}_1^T
	+
	\boldsymbol{e}_4
	\boldsymbol{e}_3^T \right)
	+
	\text{diag}(\cos \boldsymbol{\theta}_k)
	\boldsymbol{B}\left(
	\boldsymbol{e}_2
	\boldsymbol{e}_2^T
	+
	\boldsymbol{e}_4
	\boldsymbol{e}_4^T \right),
\end{align}
where $\boldsymbol{e}_1 = (1, 0,0,0 )^T$, $\boldsymbol{e}_2 = (0, 1,0,0 )^T$, $\boldsymbol{e}_3 = (0, 0,1,0 )^T$, and $\boldsymbol{e}_4 = (0, 0,0,1 )^T$, 
$\text{diag}(\sin \boldsymbol{\theta}_k) = \text{diag}(\sin \theta_k^\text{A}, \sin \theta_k^\text{B}, \sin \theta_k^\text{C}, \sin \theta_k^\text{D})$, 
$\text{diag}(\cos \boldsymbol{\theta}_k) = \text{diag}(\cos \theta_k^\text{A}, \cos \theta_k^\text{B}, \cos \theta_k^\text{C}, \cos \theta_k^\text{D})$.

When implementing quantum operations, we select the type of operations and their parameters. Accordingly, the measurement angles $\boldsymbol{\theta}_k$ are determined, and the corresponding feedforward matrix $\boldsymbol{E}_k$ is uniquely determined. Then, the other matrices $\boldsymbol{F}_k$ and $\boldsymbol{G}_k$ for the numerical feedforward are calculated using the above equations. Arbitrary displacements are also introduced, if required.

In the following sections, we will explain how the quantum operations are realized by changing the measurement angles and feedforward. After that, we will discuss two exceptional cases: readout and initialization. 

\subsection*{S1.3 Individual macronode operations}

In this section, we will describe individual macronode operations. The input computational modes $(\text{b},k)$ and $(\text{d},k)$ are transformed into the output computational modes $(\text{b},k+1)$ and $(\text{d},k+N)$. 
Prior to the measurement and the subsequent feedforward from macronode $k$, the output modes $(\text{b},k+1)$ and $(\text{d},k+N)$ are expressed as\cite{alexander2016flexible}:
\begin{align}
	\begin{pmatrix}
		\hat x_{k+1}^\text{b} \\
		\hat p_{k+1}^\text{b} \\
		\hat x_{k+N}^\text{d} \\
		\hat p_{k+N}^\text{d} \\
	\end{pmatrix}
	=
	\boldsymbol{S}(\boldsymbol{\theta}_k)
	\begin{pmatrix}
		\hat x_{k}^\text{b} \\
		\hat p_{k}^\text{b} \\
		\hat x_{k}^\text{d} \\
		\hat p_{k}^\text{d} \\
	\end{pmatrix}
	-
	\boldsymbol{T}(\boldsymbol{\theta}_k)
	\begin{pmatrix}
		\hat m_{k}^\text{A} \\
		\hat m_{k}^\text{B} \\
		\hat m_{k}^\text{C} \\
		\hat m_{k}^\text{D} \\
	\end{pmatrix}
	+\sqrt{2}
	\begin{pmatrix}
		\hat \delta_k ^\text{B} \\
		\hat \delta_k ^\text{A} \\
		\hat \delta_k ^\text{D} \\
		\hat \delta_k ^\text{C} \\
	\end{pmatrix}.
	\label{eq:OutNoF}
\end{align}
In Eq.~\eqref{eq:OutNoF}, the first term of the right-hand side represents a two-mode Gaussian operation that can be implemented by adjusting the measurement angles at macronode $k$, with the transformation matrix of the operation, $\boldsymbol{S}(\boldsymbol{\theta}_k)$. 
The second term represents an additional term that will be canceled by the measurement and feedforward displacement, as in Eq.~\eqref{eq:FFDisp}, where we set $\boldsymbol{E}_k = \boldsymbol{T}(\boldsymbol{\theta}_k)$. 
The third term represents the additional noise due to finite entanglement, i.e., non-zero nullifier. 
Here $\boldsymbol{S}(\boldsymbol{\theta}_k)$ and $\boldsymbol{T}(\boldsymbol{\theta}_k)$ are given by:
\begin{align}
	\boldsymbol{S}(\boldsymbol{\theta}_k)& = 
	(\boldsymbol{B}_0^{-1} \otimes \boldsymbol{I}_2) 
	\begin{pmatrix}
		\boldsymbol{V}(\theta_k^\text{B}+\theta_k^\text{A}, \theta_k^\text{B}-\theta_k^\text{A})  & \boldsymbol{0} \\
		\boldsymbol{0} & \boldsymbol{V}(\theta_k^\text{D}+ \theta_k^\text{C}, \theta_k^\text{D} - \theta_k^\text{C}) \\
	\end{pmatrix}
	(\boldsymbol{B}_0 \otimes \boldsymbol{I}_2), 
	\label{eq:Sthetak} \\
	\boldsymbol{T}(\boldsymbol{\theta}_k)& = 
	(\boldsymbol{B}_0^{-1} \otimes \boldsymbol{I}_2) 
	\begin{pmatrix}
		\boldsymbol{L}(\theta_k^\text{A}, \theta_k^\text{B})  & \boldsymbol{0} \\
		\boldsymbol{0} & \boldsymbol{L}(\theta_k^\text{C}, \theta_k^\text{D}) \\
	\end{pmatrix},
\end{align}
where the generalized teleportation matrix $\boldsymbol{V}$ and feedforward coefficients matrix $\boldsymbol{L}$ are defined with the rotation matrix $\boldsymbol{R}$ and squeezing matrix $\boldsymbol{\Sigma}$ as:
\begin{align}
	\boldsymbol{V}(\phi_+, \phi_-) &= 
	\boldsymbol{R}\left(\tfrac{\phi_+}{2}-\tfrac{\pi}{2} \right)
	\boldsymbol{\Sigma}\left(\tan \tfrac{\phi_-}{2}\right)
	\boldsymbol{R}\left(\tfrac{\phi_+}{2}\right) \\
	& = 
	\frac{1}{\sin\phi_-}
	\begin{pmatrix}
		\sin \phi_+ & \cos \phi_- + \cos \phi_+ \\
		\cos \phi_- - \cos \phi_+ & \sin \phi_+ \\
	\end{pmatrix}, 
	\label{eq:VbyPhi} \\
	\boldsymbol{R}(\theta) &=
	\begin{pmatrix}
		\cos \theta & -\sin \theta \\
		\sin \theta & \cos \theta \\
	\end{pmatrix},\\
	\boldsymbol{\Sigma}(t) &= 
	\begin{pmatrix}
		t & 0 \\
		0 & \tfrac{1}{t} \\
	\end{pmatrix},\\
	\boldsymbol{L}(\theta_1, \theta_2) &= 
	\frac{\sqrt{2}}{\sin(\theta_2 - \theta_1)}
	\begin{pmatrix}
		\cos \theta_2 & \cos \theta_1 \\
		\sin \theta_2 & \sin \theta_1 \\
	\end{pmatrix}.
\end{align}
Here we assume that $\theta_k^\text{B} - \theta_k^\text{A} \neq 0 \ (\mathrm{mod} \ 2\pi)$ and $\theta_k^\text{D} - \theta_k^\text{C} \neq 0 \ (\mathrm{mod} \ 2\pi)$ to avoid the divergence of $\boldsymbol{L}$. Also note that the expression of $\boldsymbol{V}$ is slightly different from that of the main manuscript.

Thus, by properly performing the measurement and feedforward, we can realize a two-mode Gaussian operation $\boldsymbol{S}(\boldsymbol{\theta}_k)$ on the input modes $(\text{b}, k)$ and $(\text{d}, k)$, resulting in the output modes $(\text{b}, k+1)$ and $(\text{d}, k+N)$ up to the additional noise due to finite entanglement.

It is also worth noting that the expression of $\boldsymbol{S}(\boldsymbol{\theta}_k)$ in Eq.~\eqref{eq:Sthetak} indicates that a quantum operation in the current system can be decomposed into two generalized teleportation as in Fig.~\ref{figS3}. Two inputs $(\text{b}, k)$ and $(\text{d}, k)$ are first combined at a 50:50 beam splitter, then the outputs are teleported and combined at another 50:50 beam splitter. 

\subsubsection*{S1.3.1 \ Output Swapping}
Before discussing specific quantum operations, we will explain an important characteristic of our system. From Eq.~\eqref{eq:VbyPhi}, it is clear that $\boldsymbol{V}(\phi_+, -\phi_-) = -\boldsymbol{V}(\phi_+, \phi_-)$. Along with the expression in Eq.~\eqref{eq:Sthetak}, it can be seen that two output modes can be swapped by exchanging $\theta_k^\text{A}$ and $\theta_k^\text{B}$:
\begin{align}
	\boldsymbol{S}(\theta_k^\text{B}, \theta_k^\text{A}, \theta_k^\text{C}, \theta_k^\text{D}) = 
	\left[
	\begin{pmatrix}
		0 & 1 \\ 1 & 0
	\end{pmatrix}
	\otimes \boldsymbol{I}_2
	\right]
	\boldsymbol{S}(\theta_k^\text{A}, \theta_k^\text{B}, \theta_k^\text{C}, \theta_k^\text{D}).
\end{align}
This property is useful when we want to apply the same operation, but want to obtain the switched outputs. 

\subsubsection*{S1.3.2 \ Single-mode operations}

\begin{figure}
	\centering
	\includegraphics[width=1\textwidth]{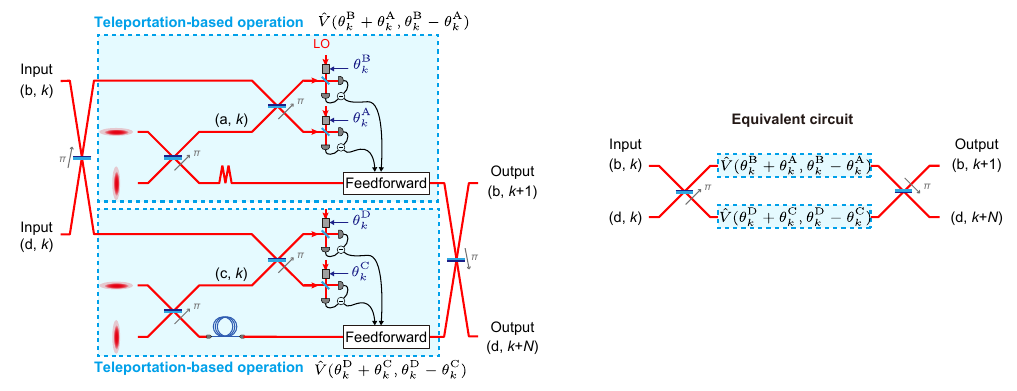}
	\caption{\textbf{Quantum operation at an individual macronode (left) and its equivalent circuit (right).} All beam splitters are 50:50 beam splitters.}
	\label{figS3}
\end{figure}

A special case of a two-mode Gaussian operation in our system is two identical but independent single-mode operations. The equivalent circuit in Fig.~\ref{figS3} shows that if we set the measurement angles as $\theta_k^\text{A} = \theta_k^\text{C}$ and $\theta_k^\text{B} = \theta_k^\text{D}$, two teleportation operations become identical and the beam splitter operations are canceled. This results in a single-mode operation applied to each input separately. In particular, we refer to this single-mode operation as \textquotedblleft crossed single-mode operation\textquotedblright. As explained in the previous section, we can swap the outputs by exchanging $\theta_k^\text{A}$ and $\theta_k^\text{B}$. Thus by setting $\theta_k^\text{B} = \theta_k^\text{C}$ and $\theta_k^\text{A} = \theta_k^\text{D}$, we can apply the single-mode operation to each input, but the outputs are swapped, which is referred to as a \textquotedblleft twisted single-mode operation\textquotedblright.

The followings are fundamental single-mode operations implementable at an individual macronode\cite{alexander2016flexible,asavanant2021time}:
\begin{itemize}
	\item \textbf{Phase rotation}
	\begin{align}
		\boldsymbol{V}\left(\psi+\tfrac{\pi}{2}, \tfrac{\pi}{2}\right) = 
		\boldsymbol{R}(\psi)=
		\begin{pmatrix}
			\cos \psi & -\sin \psi \\
			\sin \psi & \cos \psi \\
		\end{pmatrix}.
	\end{align}
	
	\item \textbf{X-invariant shear}
	\begin{align}
		\boldsymbol{V}\left(\tfrac{\pi}{2} + \tan^{-1}\kappa,\ \tfrac{\pi}{2} - \tan^{-1}\kappa\right) = 
		\boldsymbol{P}(\kappa)=
		\begin{pmatrix}
			1 & 0 \\
			2\kappa & 1 \\
		\end{pmatrix}.
	\end{align}
	
	\item \textbf{P-invariant shear}
	\begin{align}
		\boldsymbol{V}\left(\cot^{-1} \eta,\ \cot^{-1} \eta\right) = 
		\boldsymbol{Q}(\eta)=
		\begin{pmatrix}
			1 & 2\eta \\
			0 & 1 \\
		\end{pmatrix}.
	\end{align}
	
	\item \textbf{Squeezing with -90 degree phase rotation}
	\begin{align}
		\boldsymbol{V}\left(0,\ 2\tan^{-1}t\right) = 
		\boldsymbol{R}\left(-\tfrac{\pi}{2}\right)	\boldsymbol{\Sigma}(t)=
		\begin{pmatrix}
			0 & \tfrac{1}{t} \\
			-t & 0 \\
		\end{pmatrix}.
	\end{align}
	
	\item \textbf{45 degree squeezing}
	\begin{align}
		\boldsymbol{V}\left(\tfrac{\pi}{2},\ 2\tan^{-1}t\right) &= 
		\boldsymbol{R}\left(-\tfrac{\pi}{4}\right)	\boldsymbol{\Sigma}(t)\boldsymbol{R}(\tfrac{\pi}{4}) =
		\frac{1}{2}
		\begin{pmatrix}
			t+\tfrac{1}{t} & -t+\tfrac{1}{t} \\
			-t+\tfrac{1}{t} & t+\tfrac{1}{t} \\
		\end{pmatrix}
		\nonumber \\ &=
		\begin{pmatrix}
			\csc 2\psi & \cot 2\psi \\
			\cot 2\psi & \csc 2\psi \\
		\end{pmatrix}, \quad (t = \tan \psi).
	\end{align}
\end{itemize}

In addition to these fundamental single-mode operations, we can implement an arbitrary single-mode Gaussian operation using two macronodes.
\begin{itemize}
	\item \textbf{Arbitrary single-mode Gaussian operation}
	\begin{align}
		\boldsymbol{V}\left( 2\beta,\ 2\tan^{-1} e^\lambda \right)
		\boldsymbol{V}\left( \alpha - \beta,\ \tfrac{\pi}{2}\right)
		=
		\boldsymbol{R}(\alpha)\boldsymbol{\Sigma}(\lambda)\boldsymbol{R}(\beta)
	\end{align}
	
\end{itemize}

\subsubsection*{S1.3.3 \ Two-mode operations}

Next, we will show two fundamental two-mode operations, i.e., a beam splitter~\cite{alexander2016flexible} and a generalized controlled-Z operation~\cite{yoshikawa2025configurationdesignmultimodegaussian}. 

\begin{figure}
	\centering
	\includegraphics[width=0.5\textwidth]{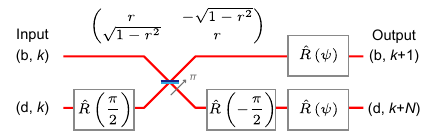}
	\caption{\textbf{Beam splitter operation at an individual macronode.}}
	\label{figS4}
\end{figure}

\begin{itemize}
	\item \textbf{Beam splitter} 
	\begin{align}
		\notag
		&\boldsymbol{S}\left(\tfrac{\psi}{2} + \tfrac{\cos^{-1}r}{2},\ \tfrac{\psi}{2} + \tfrac{\cos^{-1}r}{2}+\tfrac{\pi}{2},\ 
		\tfrac{\psi}{2} - \tfrac{\cos^{-1}r}{2},\ \tfrac{\psi}{2} - \tfrac{\cos^{-1}r}{2}+\tfrac{\pi}{2}\right) \\
		&\qquad = 
		[\boldsymbol{I}_2 \otimes \boldsymbol{R}(\psi)]
		\begin{pmatrix}
			\boldsymbol{I}_2 & \boldsymbol{0} \\
			\boldsymbol{0} & \boldsymbol{R}\left(-\tfrac{\pi}{2}\right) \\
		\end{pmatrix}
		[\boldsymbol{B}_2(r) \otimes \boldsymbol{I}_2]
		\begin{pmatrix}
			\boldsymbol{I}_2 & \boldsymbol{0} \\
			\boldsymbol{0} & \boldsymbol{R}\left(\tfrac{\pi}{2}\right) \\
		\end{pmatrix},
	\end{align}
	where $\boldsymbol{B}_2(r) = 		\begin{pmatrix}
		r & -\sqrt{1-r^2} \\
		\sqrt{1-r^2} & r \\
	\end{pmatrix}$.
	
	This equation indicates that the beam splitter operation in our system inherently introduces additional phase rotations as illustrated in Fig.~\ref{figS4}.
	
	\item \textbf{Generalized Controlled-Z}
	\begin{align}
		\boldsymbol{S}\left(\tan^{-1}\left(g-\tfrac{h}{2}\right),\ \tfrac{\pi}{2},\ \tan^{-1}\left(g+\tfrac{h}{2}\right),\ \tfrac{\pi}{2}\right) 
		=
		\begin{pmatrix}
			1  & 0  & 0  & 0 \\
			2h & 1  & g  & 0 \\
			0  & 0  & 1  & 0 \\
			g  & 0  & 2h & 1 \\		
		\end{pmatrix}.
	\end{align}
	This generalized controlled-Z operation reduces to a controlled-Z operation with $h=0$~\cite{walshe2021streamlined}.
	
\end{itemize}

\subsection*{S1.4 \ Readout and initialization on individual macronode}

In this section, we will explain two special cases, i.e., readout~\cite{alexander2016flexible} and initialization, 
where we set the measurement angles and feedforward operations differently from those used when implementing quantum operations. 

\subsubsection*{S1.4.1 \ Readout}

When performing readout at macronode $k$, we set all measurement angles to the same value $\theta_k$,
\begin{align}
	\theta_k \equiv
	\theta_k^\text{A} = 
	\theta_k^\text{B} = 
	\theta_k^\text{C} = 
	\theta_k^\text{D}.
\end{align}
In this setting, the quadrature operators $\hat{m}_k^j$ with the angle $\theta_k$ at the computational modes ($j=\text{a}, \text{b}, \text{c}, \text{d}$) and the measured modes ($j=\text{A}, \text{B}, \text{C}, \text{D}$) are related with an inverse operation of the four-splitter as, 
\begin{align}
	\begin{pmatrix}
		\hat m_{k}^\text{a} \\
		\hat m_{k}^\text{b} \\
		\hat m_{k}^\text{c} \\
		\hat m_{k}^\text{d} \\
	\end{pmatrix}
	=
	\boldsymbol{B}^{-1}
	\begin{pmatrix}
		\hat m_{k}^\text{A} \\
		\hat m_{k}^\text{B} \\
		\hat m_{k}^\text{C} \\
		\hat m_{k}^\text{D} \\
	\end{pmatrix}.
	\label{eq:ReadOut}
\end{align}

Thus, we can access the measurement outcomes of the computational modes by calculating $\boldsymbol{B}^{-1}\boldsymbol{m}_k$ with the actual measurement outcomes $\boldsymbol{m}_k = (m_k^\text{A}, m_k^\text{B}, m_k^\text{C}, m_k^\text{D})^T$.

\subsubsection*{S1.4.2 \ Initialization}

\begin{figure}
	\centering
	\includegraphics[width=0.4\textwidth]{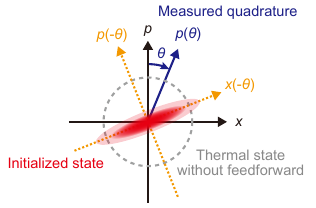}
	\caption{\textbf{Initialization.}}
	\label{figS5}
\end{figure}

Next, we consider the initialization. 
Let's assume that we are going to begin quantum operations at macronode $k$, i.e., the computational modes (b, $k$) and (d, $k$). If we simply ignore all preceding macronodes, the initial modes (b, $k$) [(d, $k$)] is a thermal state which is given by tracing out its corresponding entangled mode (a, $k-1$) [(c, $k-N$)]. Generally, a thermal state is not desirable, so we should measure the entangled mode (a, $k-1$) [(c, $k-N$)], and perform feedforward operation to prepare a purer state, i.e., a squeezed thermal state, at (b, $k$) [(d, $k$)] (see Fig.~\ref{figS5}).

In the following, for the sake of explanation, we assume that we measure macronode $k$ to prepare initial input states at (b, $k+1$) and (d, $k+N$). Also, we define the quadratures $\hat x_k^i(\theta)$ and $\hat p_i^k(\theta)$ ($i=$a, b, c, d) in the rotated frame as
\begin{align}
	\begin{pmatrix}
		\hat x_k^i (\theta) \\ \hat p_k^i(\theta)
	\end{pmatrix}
	\equiv
	\boldsymbol{R}(\theta)
	\begin{pmatrix}
		\hat x_k^i \\ \hat p_k^i
	\end{pmatrix}.
\end{align}

The EPR correlation between (a, $k$) and (b, $k+1$) [(c, $k$) and (d, $k+N$)] still exist in this rotating frame as~\cite{alexander2016flexible,asavanant2021time},
\begin{align}
	[\boldsymbol{I}_2 \otimes \boldsymbol{R}(-\theta_k)]
	\begin{pmatrix}
		\hat \delta_k^\text{B} \\ \hat \delta_k ^\text{A} \\ \hat \delta_k^\text{D} \\ \hat \delta_k ^\text{C} 
	\end{pmatrix}
	=\frac{1}{\sqrt{2}}
	\begin{pmatrix}
		- \hat x_k ^\text{a}(\theta_k) + \hat x_{k+1} ^\text{b}(-\theta_k) \\
		\phantom{-} \hat p_k ^\text{a}(\theta_k) + \hat p_{k+1} ^\text{b}(-\theta_k) \\
		- \hat x_k ^\text{c}(\theta_k) + \hat x_{k+N} ^\text{d}(-\theta_k) \\
		\phantom{-} \hat p_k ^\text{c}(\theta_k) + \hat p_{k+N} ^\text{d}(-\theta_k) \\
	\end{pmatrix}.
	\label{eq:EPRinRot}
\end{align}
Here $\delta_k^j$ ($j=\text{A}, \text{B}, \text{C}, \text{D}$) is a nullifier given in Eq.~\eqref{eq:HD_nullifier}.

When we measure macronode $k$, we set all the measurement angles to the same value $\theta_k$ as in the case of readout. Thus, according to Eq.~\eqref{eq:ReadOut}, we can access the quadratures $\hat m_{k}^{i}$ ($i=$a, b, c, d) which is equivalent to $\hat p_{k}^{i} (\theta_k)$.

Eq.~\eqref{eq:EPRinRot} can be rearranged by replacing $\hat p_{k}^{i} (\theta_k)$ with $\hat m_{k}^{i}$ as,
\begin{align}
	\begin{pmatrix}
		\hat x_{k+1}^\text{b} (-\theta_k) \\
		\hat p_{k+1}^\text{b} (-\theta_k) \\
		\hat x_{k+N}^\text{d} (-\theta_k) \\
		\hat p_{k+N}^\text{d} (-\theta_k) \\
	\end{pmatrix}
	=
	\begin{pmatrix}
		\hat x_k^\text{a}(\theta_k) \\
		0 \\
		\hat x_k^\text{c}(\theta_k) \\
		0 \\
	\end{pmatrix}
	-
	\begin{pmatrix}
		0 \\
		\hat m_k^\text{a} \\
		0 \\
		\hat m_k^\text{c} \\
	\end{pmatrix}
	+\sqrt{2}
	[\boldsymbol{I}_2 \otimes \boldsymbol{R}(-\theta_k)]
	\begin{pmatrix}
		\hat \delta_k^\text{B} \\ \hat \delta_k ^\text{A} \\ \hat \delta_k^\text{D} \\ \hat \delta_k ^\text{C} 
	\end{pmatrix}.
\end{align}
The second term of the right-hand side of the equation can be canceled out by the feedforward displacement to macronodes $k+1$ and $k+N$ as in Eq.~\eqref{eq:FFDisp}. The feedforward coefficient matrix $\boldsymbol{E}_k$ is given as
\begin{align}
	\boldsymbol{E}_k = \left[ \begin{pmatrix} -\sin\theta & 0 \\ \cos\theta & 0 \end{pmatrix}  \otimes \boldsymbol{I}_2\right]
	\boldsymbol{B}^{-1}.
\end{align}

After the feedforward operation, the initialized modes at (b, $k+1$) and (d, $k+N$) are squeezed thermal states with the squeezing angle $-\theta_k$, whose quadrature means and mean squares are given as,
\begin{align}
	&\big\langle \hat x_{k+1}^\text{b} (-\theta_k) \big\rangle = 
	\big\langle \hat p_{k+1}^\text{b} (-\theta_k) \big\rangle = 
	\big\langle \hat x_{k+N}^\text{b} (-\theta_k) \big\rangle = 
	\big\langle \hat p_{k+N}^\text{b} (-\theta_k) \big\rangle = 0, \\
	&\big\langle \big(\hat x_{k+1}^\text{b} (-\theta_k) \big)^2\big\rangle = \frac{\cos^2\theta_k }{4}\left(\text{e}^{2r_\text{A}} + \text{e}^{-2r_\text{B}}  \right) + \frac{\sin^2\theta_k }{4}\left(\text{e}^{-2r_\text{A}} + \text{e}^{2r_\text{B}}  \right) , \\
	&\big\langle \big(\hat p_{k+1}^\text{b} (-\theta_k) \big)^2\big\rangle = 
	\cos^2\theta_k \text{e}^{-2r_\text{A}} + \sin^2\theta_k\text{e}^{-2r_\text{B}},\\ 
	&\big\langle \big(\hat x_{k+N}^\text{d} (-\theta_k) \big)^2\big\rangle = \frac{\cos^2\theta_k }{4}\left(\text{e}^{2r_\text{C}} + \text{e}^{-2r_\text{D}}  \right) + \frac{\sin^2\theta_k }{4}\left(\text{e}^{-2r_\text{C}} + \text{e}^{2r_\text{D}}  \right) , \\
	&\big\langle \big(\hat p_{k+N}^\text{d} (-\theta_k)  \big)^2\big\rangle = 
	\cos^2\theta_k \text{e}^{-2r_\text{C}} + \sin^2\theta_k\text{e}^{-2r_\text{D}}.
\end{align}

It is also noted that we can add an arbitrary displacement as in Eq.~\eqref{eq:NumFF}. Thus, initial states that can be prepared are displaced squeezed thermal states.

\subsection*{S1.5 \ External input coupling and implementation of non-Gaussian operations}

\begin{figure} 
	\centering
	\includegraphics[width=1\textwidth]{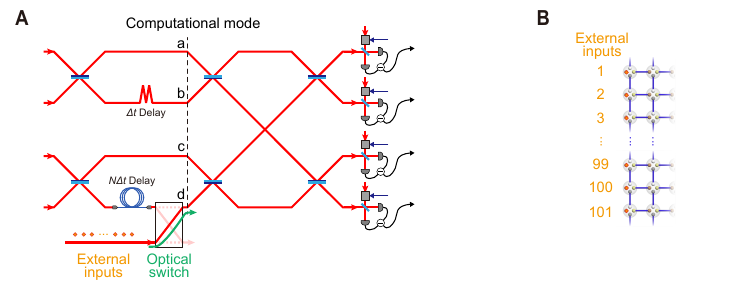}
	
	\caption{\textbf{External input coupling.} ({\bf A}) Schematic of the external input coupling. An optical switch is used to inject external quantum states into the computational modes d. ({\bf B}) Graph representation of the injected external inputs.}
	\label{figS6} 
\end{figure}

To incorporate external quantum states into the platform, an optical switch is used to replace the quantum state in the computational modes d with an external input mode (Fig.~\ref{figS6}A). By synchronizing the switching timing with the system clock, external quantum states—such as non-Gaussian states or any states other than those prepared during the initialization process—are coupled into the computational modes d. In the graph representation, these coupled states are treated as external inputs assigned to specific macronodes (Fig.~\ref{figS6}B). While the current demonstration in the main text uses Gaussian states prepared by initialization, external input coupling significantly broadens the computational capabilities. 
Specifically, combined with the programmable routing function explained in the main text, this would allow for the flexible integration of probabilistically generated non-Gaussian states into the target modes. Such a capability is a prerequisite for functioning as a processor that can effectively manage non-Gaussian resources.

\begin{figure} 
	\centering
	\includegraphics[width=1\textwidth]{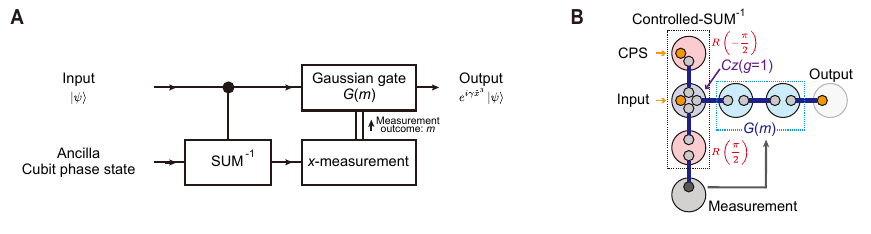}
	
	\caption{\textbf{Implementation of non-Gaussian operations.} ({\bf A}) Conceptual circuit diagram for a cubic phase gate utilizing a non-Gaussian ancillary state and non-linear feedforward. ({\bf B}) Mapping of the non-Gaussian operation implementation onto the graph. CPS: Cubic phase state.}
	\label{figS7} 
\end{figure}

The implementation of universal quantum computation requires the execution of non-Gaussian operations, such as the cubic phase gate. In a measurement-based architecture, these operations are typically realized by utilizing a non-Gaussian ancillary state, such as a cubic phase state~(Fig.~\ref{figS7}A) \cite{gottesman2001encoding}. Specifically, the input state and the cubic phase state are made to interact via a SUM$^{-1}$ gate, followed by an $x$-measurement on one of the modes. A Gaussian operation is then applied to the remaining mode conditioned on the measurement result to complete the cubic phase gate. Our platform is designed to map this process onto the graph structure (Fig.~\ref{figS7}B), where the ancillary state is introduced and subsequently manipulated through non-linear feedforward~\cite{sakaguchi2023nonlinear}.

\subsection*{S1.6 \ Estimation of transformation matrices}

In this section, we will explain how to estimate a transformation matrix. We assume that $n$ input modes with quadratures $\boldsymbol{\hat q_\text{in}}$ are transformed into $n$ output modes with quadratures $\boldsymbol{\hat q_\text{out}}$ via a transformation matrix $\boldsymbol{S}$ and an additional noise term $\boldsymbol{\hat \Delta}_{\text{noise}}$,
\begin{align}
	\boldsymbol{\hat q_\text{out}} = \boldsymbol{S} 
	\boldsymbol{\hat q_\text{in}} + \boldsymbol{\hat \Delta}_{\text{noise}}.
	\label{eq:IndivSmatSup}
\end{align}
While two input and two output modes are considered when estimating the transformation matrix of an individual macronode operation, an arbitrary number of input and output modes up to 101 can be considered in general.

 To estimate the transformation matrices, we do not apply initialization operations. This ensures that each input mode maintains an entangled mode, which is a reference mode measurable without disturbing the operations. 
 
 In advance, we measure the quadrature correlations between the reference and input modes 
 $\left \langle \boldsymbol{\hat q}_\text{in} \boldsymbol{\hat q}_\text{ref}^T \right \rangle$
 where $\left\langle \cdot \right\rangle$ means an average of symmetric ordered operators. It should be noted that this 
 $\left \langle \boldsymbol{\hat q}_\text{in} \boldsymbol{\hat q}_\text{ref}^T \right \rangle$
 is a diagonal matrix. 
 Then, we measure the correlation between the reference and output modes 
 $\left \langle \boldsymbol{\hat q}_\text{out} \boldsymbol{\hat q}_\text{ref}^T \right \rangle$. 
 Since the additional noise term 
$\boldsymbol{\hat \Delta}_\text{noise}$ 
 is independent from the reference mode quadratures, i.e., 
$\left \langle \boldsymbol{\hat \Delta}_\text{noise} \boldsymbol{\hat q}_\text{ref}^T \right \rangle=0$, 
 the transformation matrix in Eq.~\eqref{eq:IndivSmatSup} is given as
\begin{align}
 \boldsymbol{S} = \left \langle \boldsymbol{\hat q}_\text{out} \boldsymbol{\hat q}_\text{ref}^T \right \rangle \left \langle \boldsymbol{\hat q}_\text{in} \boldsymbol{\hat q}_\text{ref}^T \right \rangle^{-1}.
\end{align}
Thus, this method allows us to estimate the transformation matrices via quadrature correlations.

\subsection*{S1.7 \ Experimental setup}
In this section, we will explain additional information about our experimental setup that are not covered in the main manuscript.

In our experiment, the laser source is a continuous-wave fiber laser at 1545.32~nm.
The four OPAs, pumped by 772.66~nm laser beams, produce four squeezed vacuum states with around 6~THz bandwidth \cite{kashiwazaki2021fabrication,kashiwazaki2023over}. The squeezed vacuum state is assumed to be packetized in a $\Delta t = 10~\text{ns}$ duration, which is primarily limited by the electronic devices such as a homodyne detector and an FPGA. The corresponding length of the wavepacket in free space is 3~m. While an additional 3~m free-space optical path is added for the short delay line (see Fig.~\ref{figS1}), 200~m of fiber is used for the long delay line as $N$ is set to 101. The lengths of the delay lines are finely tuned using free-space translation stages.

The interference visibilities at the homodyne detectors are $98\% \sim 99\%$ on average. Total optical propagation losses are estimated as less than $1\%$ for the path including the short delay line, and less than $8\%$ for the path including the long delay line. LO powers at the homodyne detectors are set to around 2.7~mW, giving a shot-noise clearance of about 17~dB at 100~MHz.

For a stable operation, the experimental setup is located inside a temperature-controlled booth. In addition, the optics are covered to avoid vibration and air flow. The optical path lengths are feedback-controlled by FPGA controllers. For this purpose, frequency-shifted probe beams are injected into the OPAs, and a few $\mu$W of probe beams propagate in the interferometer. To prevent the probe beams from affecting the measurements of quantum states at the homodyne detectors, we switch off the probe beams when measuring the quantum states and performing quantum operations. The switching period is 1.7~ms, with around 1~ms measurement window. During this measurement window, feedback signals to actuators are held, and the optical path length slightly drifts due to environmental fluctuations, which can be seen in the nullifier measurements in next section.

The quadrature signals from the homodyne detectors, $\hat q^{j}(t) = \hat x ^ j (t) \sin \theta^j(t) + \hat p^j(t) \cos \theta^j(t)$ ($j=\text{A}, \text{B}, \text{C}, \text{D}$), are recorded by an FPGA (AMD ZCU208) with an ADC sampling frequency of 1.6~GHz. 16 points of the quadrature signals are used to calculate a quadrature value of each wavepacket, $q_k^j=\int dt f(t) \hat q^j(t)$, where $f(t)=t\exp (-\gamma^2t^2)$ with $\gamma = 2\pi \times 140~\text{MHz} $. 
The detailed method of packetization can be found in, for example, the supplementary information of Refs~\cite{yokoyama2013ultra,asavanant2019generation}. 

The same FPGA is used to generate signals that drive electro-optic modulators for modulating LO phases, $\theta^j(t)$, at the homodyne detectors. The sampling frequency of the FPGA's DAC is 3.2~GHz, synchronized with the FPGA's ADC. The modulation signal for each wavepacket has a duration of 10~ns with the first and last 2.5~ns reversed to produce DC-canceled signals.

\section*{S2 \ Additional experimental data}

In this section, we will explain additional experimental data.

\subsection*{S2.1 \ Nullifier}

First, we present the measurement results of nullifiers in Eq.~\eqref{eq:HD_nullifier}.
The homodyne angles are set to $\pi/2$ to measure the $x$-quadratures at all homodyne detectors, or to zero to measure the $p$-quadratures. The outputs are recorded and packetized in the FPGA. A linear combination of the packetized quadratures is calculated to produce each nullifier $\hat \delta^j_k$ ($j=$A, B, C, D)~\cite{yokoyama2013ultra,yoshikawa2016invited,asavanant2019generation}. 

\begin{figure}
	\centering
	\includegraphics[width=1\textwidth]{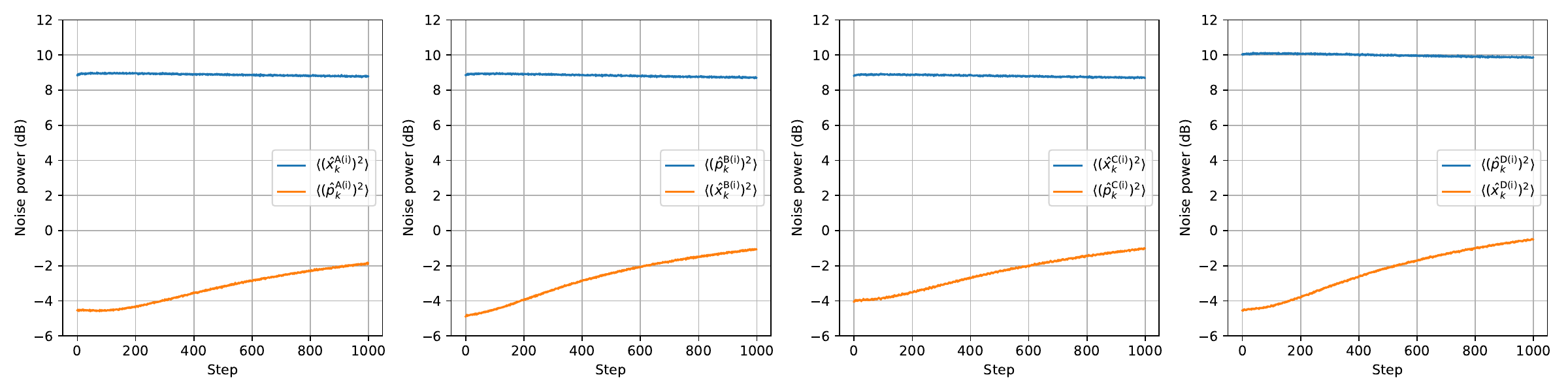}
	\caption{\textbf{Nullifier measurement.} In addition to nullifiers $\hat \delta^j_k = p^\text{A(i)}_k, x^\text{B(i)}_k, p^\text{C(i)}_k, x^\text{D(i)}_k$, the corresponding anti-squeezing levels ($x^\text{A(i)}_k, p^\text{B(i)}_k, x^\text{C(i)}_k, p^\text{D(i)}_k$) are shown for comparison. }
	\label{figS8}
\end{figure}

To characterize the nullifiers during a sequence of quantum operations, we average the variances over one full turn of the helical structure of the entanglement, i.e., 
$ V^j_m = \Sigma_{k=1}^{N} \langle ( \hat \delta^j_{mN+k} )^2 \rangle / N$.
These nullifiers roughly correspond to the resource EPR states at the $m$-th step of parallel quantum teleportation in the main manuscript. 
Figure~\ref{figS8} shows the variances (noise powers) of the nullifiers as a function of step $m$. 
Each data point is averaged over 1,000 measurement repetitions. Since the feedback control of the optical path lengths is switched on and off every 1.7~ms, the nullifiers degrade during the measurement window. This degradation affects the quality of teleportation as seen in the results of parallel teleportation. However, after multiple teleportation operations, the quantum state becomes noisy, and the slight depletion of entanglement has little impact on it.

\begin{figure}
	\centering
	\includegraphics[width=1\textwidth]{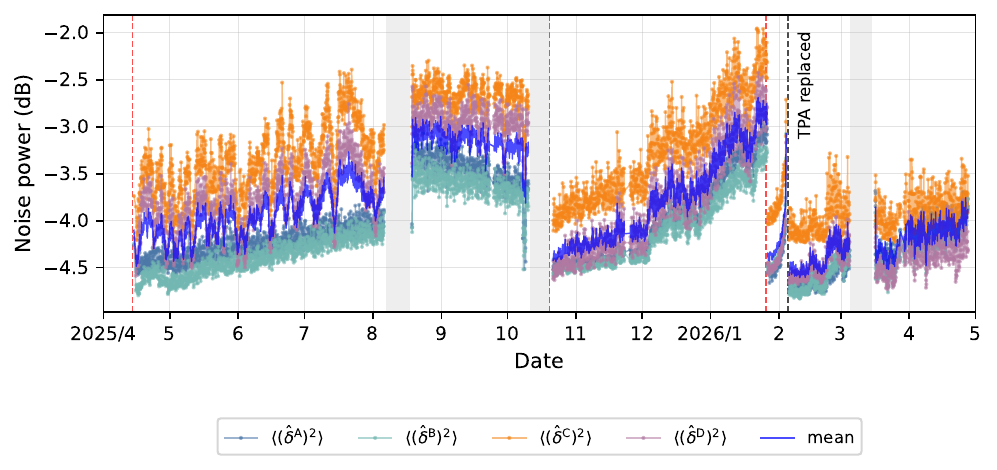}
	\caption{\textbf{Long-time nullifier measurements.} }
	\label{figS9}
\end{figure}

In addition to characterization of the nullifier during computations, we have also examined and recorded the long-term stability of the systems. Figure~\ref{figS9} shows the nullifier measurements over one year of operation. The first 30 steps of the nullifiers were measured and averaged, with measurements repeated every two hours. The orange dashed lines indicate when the optical mirrors were aligned. Over the course of the year, we performed alignment only three times, yet the system's nullifiers remained more or less stable. The shaded areas indicate maintenance and upgrade periods. The black dashed line marks when the tapered amplifier for 773~nm broken and was replaced. Overall, our system demonstrates high stability and ease of operation, requiring a few adjustments over a year.

\subsection*{S2.2 \ Individual macronode operations}

\begin{figure}
	\centering
	\includegraphics[width=1\textwidth]{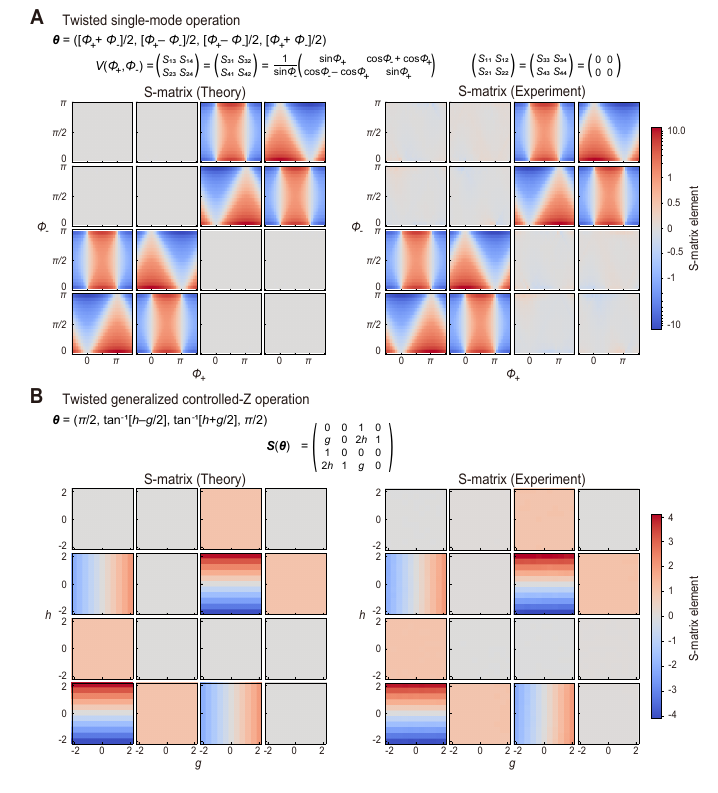}
	\caption{\textbf{Individual macronode operations.} (\textbf{A}) Twisted single-mode operation. (\textbf{B}) Twisted generalized controlled-Z operation.}
	\label{figS10}
\end{figure}

\begin{figure}
	\centering
	\includegraphics[width=1\textwidth]{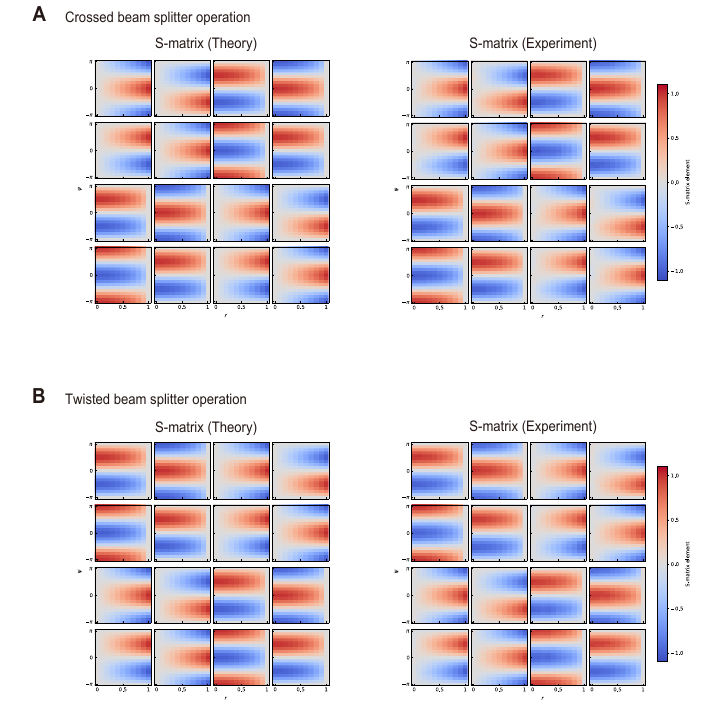}
	\caption{\textbf{Beam splitter operations.} (\textbf{A}) Crossed beam splitter operation. (\textbf{B}) Twisted beam splitter operation.}
	\label{figS11}
\end{figure}

\begin{figure}
	\centering
	\includegraphics[width=0.9\textwidth]{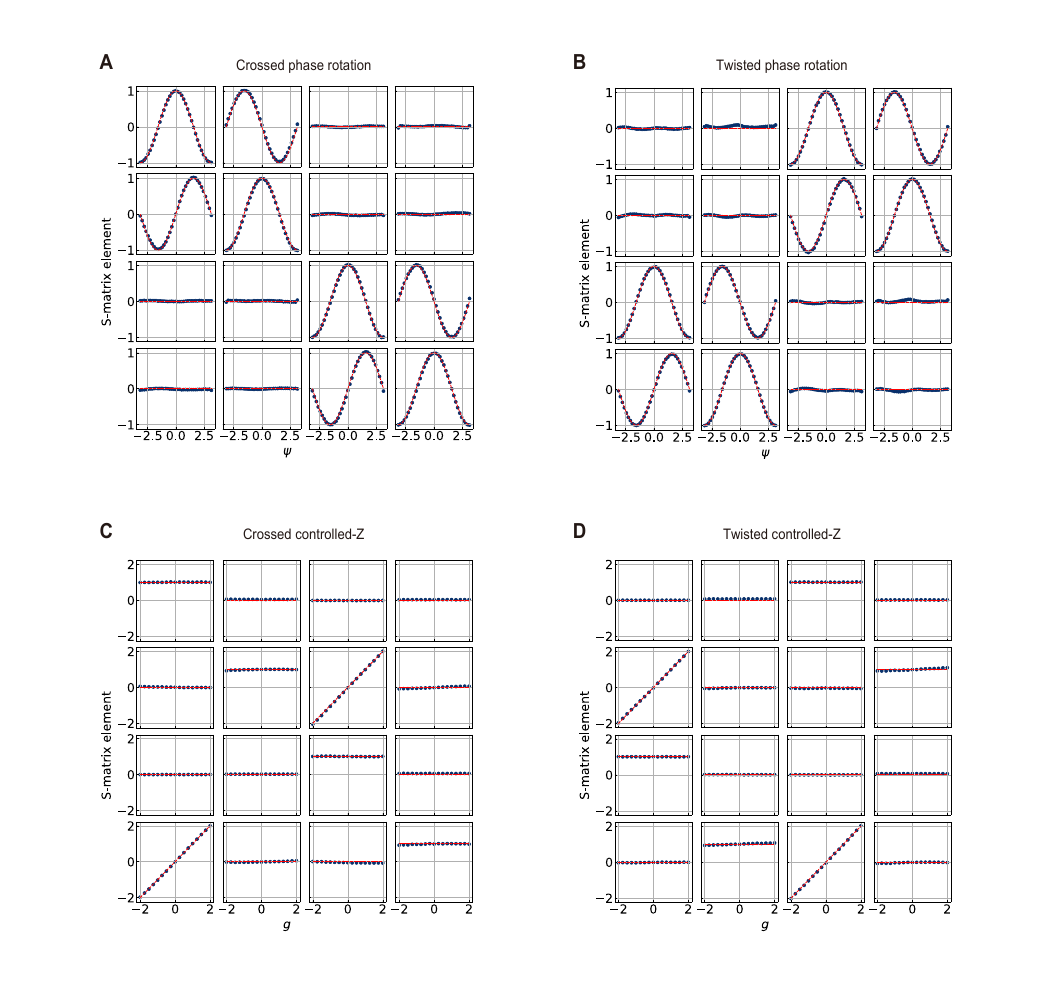}
	\caption{\textbf{Phase rotation and controlled-Z operations.} Dotted lines are the theoretical traces. (\textbf{A}) Crossed phase rotation. (\textbf{B}) Twisted phase rotation. (\textbf{C}) Crossed controlled-Z operation. (\textbf{D}) Twisted controlled-Z operation.}
	\label{figS12}
\end{figure}

\begin{figure}
	\centering
	\includegraphics[width=0.9\textwidth]{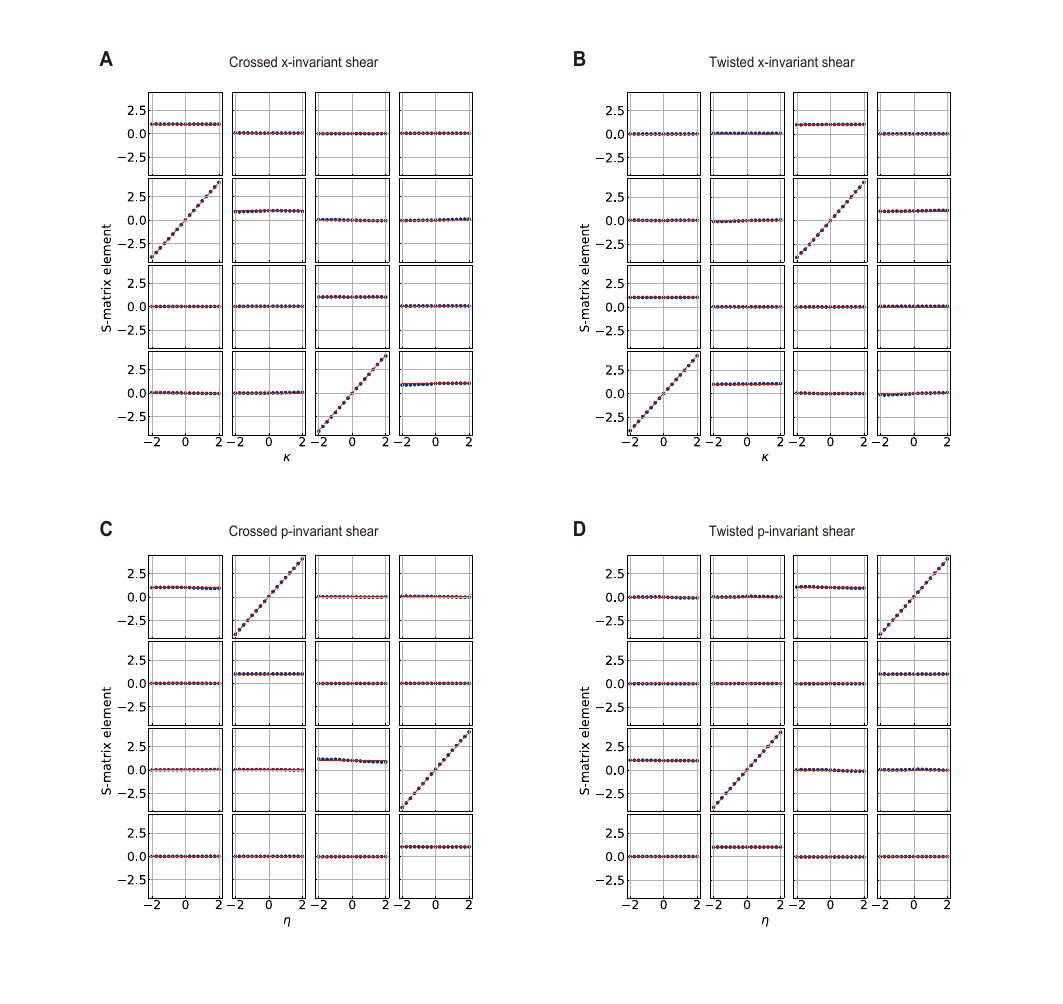}
	\caption{\textbf{Shear operations.} Dotted lines are the theoretical traces. (\textbf{A}) Crossed $x$-invariant shear. (\textbf{B}) Twisted $x$-invariant shear. (\textbf{C}) Crossed $p$-invariant shear. (\textbf{D}) Twisted $p$-invariant shear.}
	\label{figS13}
\end{figure}

\begin{figure}
	\centering
	\includegraphics[width=1\textwidth]{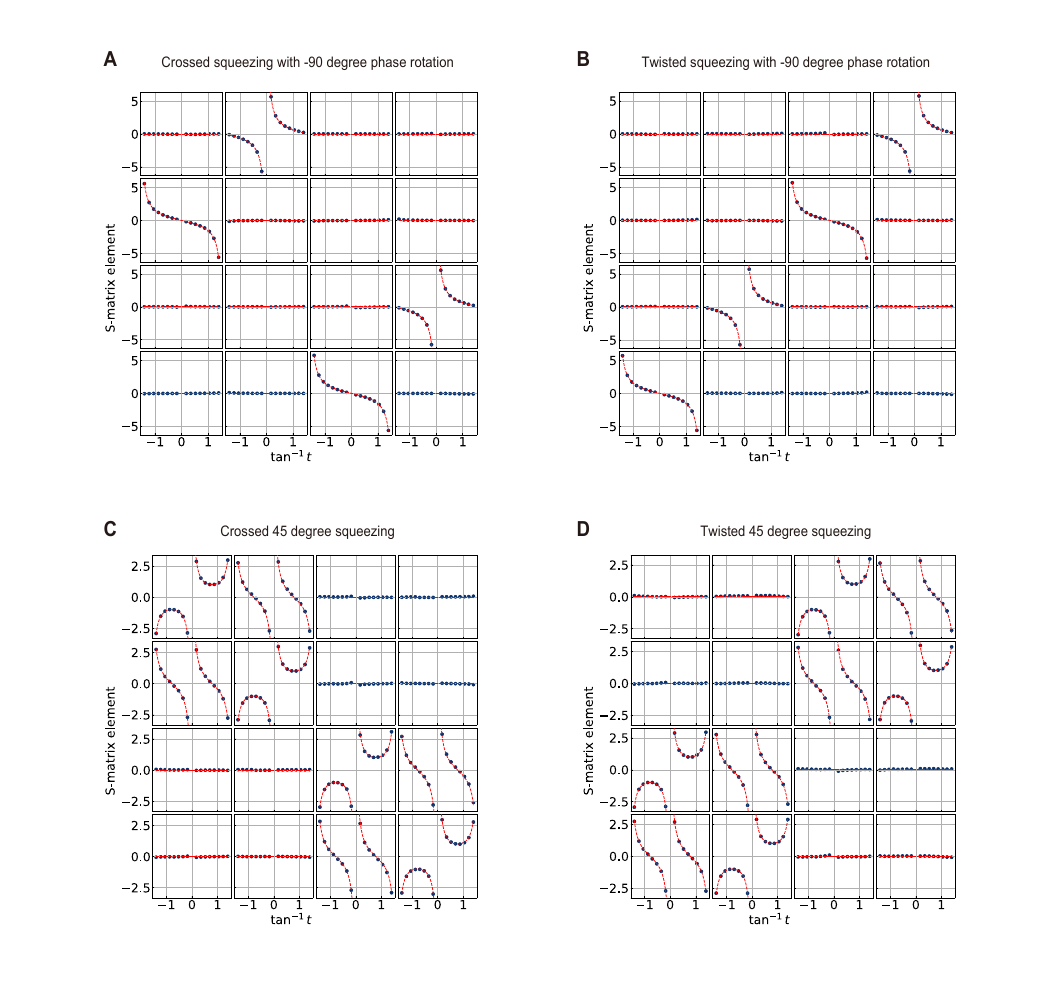}
	\caption{\textbf{Squeezing operations.} Dotted lines are the theoretical traces. (\textbf{A}) Crossed squeezing with $-90$ degree phase rotation. (\textbf{B}) Twisted squeezing with $-90$ degree phase rotation. (\textbf{C}) Crossed 45 degree squeezing. (\textbf{D}) Twisted 45 degree squeezing.}
	\label{figS14}
\end{figure}

In this section, we will show additional results for the reconstructed transformation matrices $S$ for individual macronode operations. Figure~\ref{figS10} shows twisted single-mode operations and twisted generalized controlled-Z operations. Figure~\ref{figS11} shows both crossed and twisted beam splitter operations. Figure~\ref{figS12} shows phase rotation and controlled-Z operations. Figure~\ref{figS13} and Figure~\ref{figS14} show the shear and squeezing operations, respectively.

\begin{figure}
	\centering
	\includegraphics[width=1\textwidth]{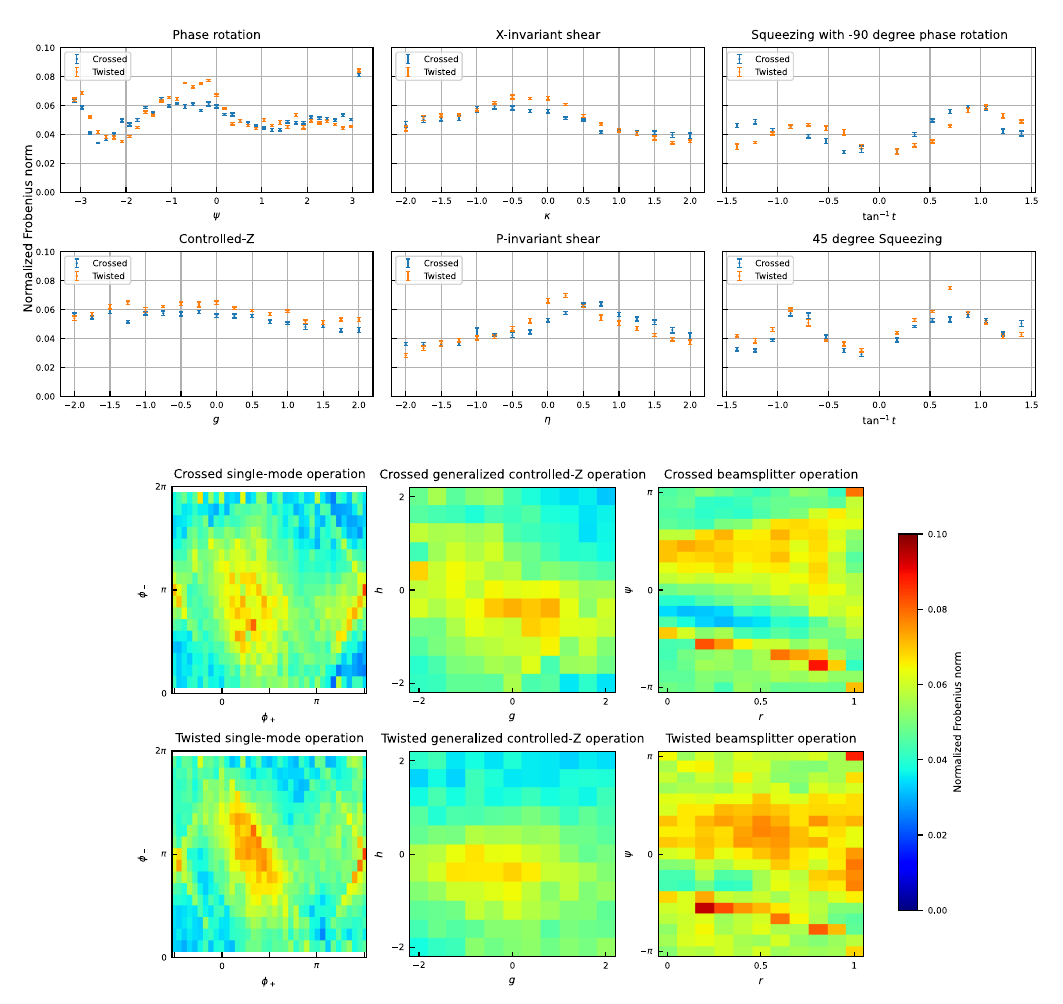}
	\caption{\textbf{Normalized Frobenius norm of the error for operations.}}
	\label{figS15}
\end{figure}

Figure~\ref{figS15} shows the normalized Frobenius norm of the error for each operation as a function of parameters.

\subsection*{S2.3 \ Parallel teleportation with classical benchmark}

\begin{figure}
	\centering
	\includegraphics[width=0.4\textwidth]{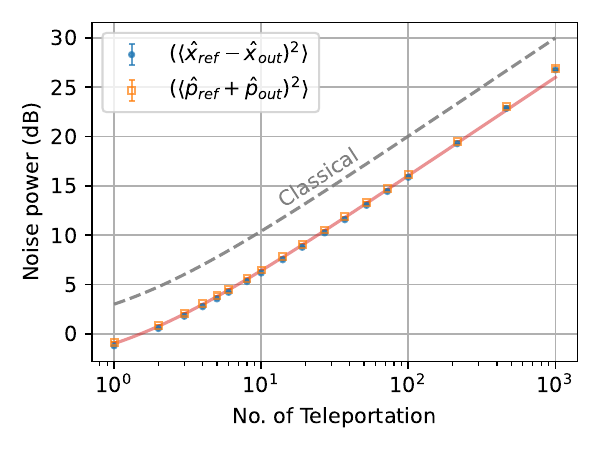}
	\caption{\textbf{Parallel teleportation with classical benchmark.} The dotted line is the classical benchmark, which is the noise level obtained by an equivalent process performed without squeezing (i.e., using only vacuum states). The solid line is the theoretical prediction with the estimated nullifiers.}
	\label{figS16}
\end{figure}

Figure~\ref{figS16} shows the results of parallel teleportation with the classical benchmark of sequential teleportation. The data is the same as in the main manuscript, but the mean and standard deviation are calculated for each step using the results of 101 parallel teleportation.
The classical benchmark is calculated as teleportation is performed without entanglement. The experimental results are well below the classical benchmark over all number of teleportation.

\clearpage

\subsection*{S2.4 \ Helical teleportation}

\begin{figure}
	\centering
	\includegraphics[width=1\textwidth]{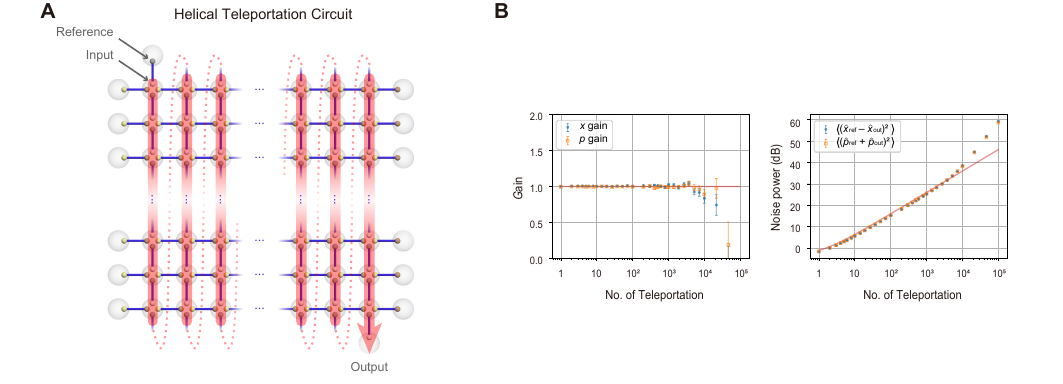}
	\caption{\textbf{Helical teleportation.} (\textbf{A}) Graph representation. (\textbf{B}) The gains of teleportation (left) and the noise power of teleportation (right).}
	\label{figS17}
\end{figure}

In addition to the \textquotedblleft parallel\textquotedblright~teleportation, we implement \textquotedblleft helical\textquotedblright~teleportation as shown in Fig.~\ref{figS17}A. Another input mode at the first macronode is teleported along with the helical axis of the graph, achieving up to 100,000 steps teleportation. Figure~\ref{figS17}B shows the results of the helical teleportation, where the teleportation is repeated by 1,000,000 times. 
The gains are around unity as it is teleportation, but the uncertainty of the gain increases at around 1,000 times and more teleportation. In particular, the gains of 10,000 times and more teleportation show large errors. This is partly because the large noise at the large number of teleportation makes the accurate estimation of the amplitude of outputs impractical, and also there may be small offsets or rounding errors accumulated over a large number of teleportation. Regarding the noise power, they agree well with theory up to around 10,000 times of teleportation. The deviations of the noise power at the large number of teleportation come from the depleted entanglement due to the locking methods as seen in the nullifier measurements.

\subsection*{S2.5 \ Programmable quantum state routing circuits}
Figures \ref{figS18} and \ref{figS19} are graph representations of the 101 modes quantum state routing circuits in ascending, and descending order, respectively.


\clearpage

\begin{figure}
	\centering
	\includegraphics[height=0.95\textheight]{FigS18.pdf}
	\caption{\textbf{Graph designed for 101 modes quantum state routing in ascending order.}}
	\label{figS18}
\end{figure}

\begin{figure}
	\centering
	\includegraphics[height=0.95\textheight]{FigS19.pdf}
	\caption{\textbf{Graph designed for 101 modes quantum state routing in descending order.}}
	\label{figS19}
\end{figure}



\end{document}